\documentclass[apj]{emulateapj}  

\catcode`\"=\active\let"=\"  

\def\3{{\ss} }

\def\c12{{1\over 2}}

\def\plusplus{\raise 0.3ex\hbox{${\scriptstyle ++}$}{}}

\shorttitle{Tidal evolution of dSph galaxies}  
\shortauthors{Pe\~{n}arrubia,  Navarro \& McConnachie}  
\slugcomment{Submitted to ApJ}  
  
\newcommand{\oversim}[2]{\protect{\mbox{\lower0.5ex\vbox{%
   \baselineskip=0pt\lineskip=0.2ex   
   \ialign{$\mathsurround=0pt #1\hfil##\hfil$\crcr#2\crcr\sim\crcr}}}}}    
\begin{document}   
   
\title[Tidal evolution of dSph galaxies]{The Tidal Evolution of Local   
Group Dwarf Spheroidals}   
   
\author{Jorge Pe\~{n}arrubia\altaffilmark{1}, Julio F. Navarro\altaffilmark{2} \&   
Alan W. McConnachie}    
   
 \affil{Department of Physics and Astronomy, University of Victoria,   
3800 Finnerty Rd., Victoria, BC, V8P 5C2, Canada}   
   
\altaffiltext{1}{Email: jorpega@uvic.ca}   
\altaffiltext{2}{Fellow of the Canadian Institute for Advanced Research}   
   
\begin{abstract}   
We use N-body simulations to study the evolution of dwarf spheroidal  
galaxies (dSphs) driven by galactic tides. We adopt a  
cosmologically-motivated model where dSphs are dark matter-dominated  
systems on eccentric orbits whose stellar component may be  
approximated by a King model embedded within an NFW halo.  We find  
that these NFW-embedded King models are extraordinarily resilient to  
tides; the density profile of the stellar component still resembles a  
King model even after losing more than 99\% of the stars. As tides  
strip the galaxy, the stellar luminosity, $L$, velocity dispersion,  
$\sigma_0$, central surface brightness, $\Sigma_0$, and core radius,  
$R_c$, decrease monotonically.  Remarkably, we find that the evolution  
of these parameters is solely controlled by the total amount of mass  
lost from within the luminous radius. Of all parameters, the core  
radius is the least affected: after losing 99\% of the stars, $R_c$  
decreases by just a factor of $\sim 2$, implying that even in the  
event of extreme mass loss the core radius is a robust measure of the  
original size of the system. Contrary to naive expectations, tides  
tend to make dSphs {\it more dark-matter dominated}. This is because  
the tightly bound central dark matter ``cusp'' is more resilient to  
disruption than the comparatively more loosely bound ``cored'' King  
profile. We examine whether tidal effects may help to explain the  
extremely large mass-to-light ratios of some of the newly-discovered  
ultra-faint Milky Way dwarfs as tidal remnants of once brighter  
systems. Although dSph tidal evolutionary tracks parallel the observed  
scaling relations in the luminosity-radius plane, they predict too  
steep a change in velocity dispersion compared with the observational  
estimates hitherto reported in the literature. The ultra-faint dwarfs  
are thus unlikely to be the tidal remnants of systems like Fornax,  
Draco, or Sagittarius. Despite spanning four decades in luminosity,  
dSphs appear to inhabit halos of comparable peak circular velocity,  
lending support to scenarios that envision dwarf spheroidals as able  
to form only in halos above a certain mass threshold.  
\end{abstract}   
    
\section{Introduction}   
\label{sec:intro}   
   
Dwarf spheroidal galaxies (dSphs) are the faintest objects believed to  
be dominated by dark matter, and as such play a critical role in  
galaxy formation models. Because of their extremely low luminosity,  
this population of dwarf galaxies has been studied in detail almost  
exclusively in the Local Group (LG). The prototype dSph is a low  
surface brightness spheroid of old stars, with little detectable gas  
and few signs of ongoing or very recent star formation. They have  
luminosities and velocity dispersions comparable to globular clusters,  
but their much larger size denote the presence of large amounts of  
dark matter and set them apart from star clusters.  
  
The best studied (``classical'') Local Group dSphs span roughly two  
decades in luminosity, $3\times 10^5 \lesssim L/L_{\odot} \lesssim  
3\times 10^7$, from the brightest (Sagittarius) to the faintest  
(Draco). In contrast to the large range spanned in $L$, all of these  
systems have similar velocity dispersion $\sigma \sim 7$--$10$ km/s,  
and share a characteristic value for the dynamically-inferred mass  
within the luminous radius, of order $\sim 10^7 M_\odot$ (for a  
review, see, e.g., Mateo 1998).  
   
This characteristic mass scale is interesting from a cosmological  
perspective (see, e.g., Gilmore et al 2007 for a recent  
discussion). Indeed, in the current $\Lambda$CDM paradigm of structure  
formation, dark matter halos of such mass outnumber the known dSphs by  
a factor of 10 to 100, a result that highlights the low formation  
efficiency of visible galaxies in low mass halos (Kauffmann, White \&  
Guiderdoni 1993, Cole et al 1994, Klypin et al 1999, Moore et al  
1999). A number of mechanisms have been proposed to explain the  
scarcity of sub-luminous galaxies in the $\Lambda$CDM scenario; these  
typically appeal to feedback effects associated with stellar  
evolution, as well as to heating from external UV radiation in order  
to hinder the cooling of gas and its transformation of stars in  
low-mass halos (Efstathiou 1992, Bullock et al 2000, Somerville 2002,  
Benson et al 2002).  
   
Although these models can, with plausible assumptions, reconcile the  
faint end of the galaxy luminosity function with the low-mass end of  
the dark matter halo mass function, they have so far provided little  
insight into the origin of the structural properties of dSphs. For  
example, the large variation in mass-to-light ratio between dSphs  
noted above (Mateo 1998); the well-defined ``King-like'' core-halo  
structure of the stars (e.g. Irwin \& Hatzidimitriou 1995); the  
intriguing presence of multiple dynamical stellar components (Tolstoy  
et al. 2004; Battaglia et al. 2006; Ibata et al. 2006);   
the near constancy of their velocity dispersion profiles  
(e.g. Wilkinson et al. 2004; Mu\~noz et al. 2005; Walker et al. 2006, 2007;  
Koch et al. 2007a,b; Mateo et al. 2007)---all of these are well established observational  
facts of unclear origin and so far lacking definitive theoretical  
underpinning.  
   
Clues to these questions may be gathered from the spatial distribution  
of dSphs, which tend to cluster tightly around the giant spirals of  
the Local Group. This observation is a basic ingredient of dSph  
formation models, many of which rely on proximity to a giant galaxy in  
order to shield them from accretion of intergalactic material that  
might otherwise fuel recent episodes of star formation. Besides  
depriving them from star formation fuel, proximity to a central galaxy  
may allow for tidal interactions to sculpt spheroidals out of  
otherwise irregular galaxies, as in the model championed by Mayer et  
al (2007, see additional references therein).  
   
One oddity in this context is the presence of two {\it isolated} dSphs,  
Cetus and Tucana, which are found far from both M31 and MW, and  
orbiting in the outskirts of the Local Group. A possible resolution,  
however, has been recently argued by Sales et al (2007), who point out  
that in hierarchical models satellite galaxies often accrete into the  
primary as systems of multiple objects. The multiple-body interactions  
involved in the tidal dissolution of these systems may propel some  
satellites onto highly-energetic orbits that would take them far away  
from their parent galaxy. More difficult to reconcile with this  
scenario is the existence of dSphs infalling for the first time into  
the Local Group, as has been argued for And XII (Martin et al. 2006,  
Chapman et al. 2007). If correct, this interpretation would imply that  
at least some dSphs are able to form in a radically different  
environment well away from giant spirals.  
   
The role of tides in shaping dSphs is nowhere clearer than in the case  
of the Sagittarius dwarf (Ibata et al 1994), which is in the process  
of shedding, as a result of its interaction with the Milky Way, a  
significant fraction of its stars in large ``tails'' that have been  
detected all across the sky (Mart\'inez-Delgado et al. 2001, Majewski et al. 2003; Belokurov et  
al. 2006). Tides are likely to continue to strip Sagittarius in the  
future, and it is therefore a fair and interesting question to ponder  
what the eventual fate of Sagittarius will be once it loses most of  
its stars.  
   
This is a particularly timely issue in light of the newly discovered  
population of ultra-faint dwarfs around M31 and MW. These systems are  
roughly 10 to 100 times fainter than the ``classical'' dSphs  
introduced above; have amorphous morphology and very low surface  
brightness (Belokurov et al. 2007, Zucker et al. 2007 and references  
therein). Could these ultra-faint dwarfs be the tidal remnants of a  
population of brighter objects now in the process of being tidally  
stripped into oblivion?   
  
Signs of ongoing tidal disruption would be much harder to detect in  
systems fainter than Sagittarius (the brightest of all known dSphs),  
but such a possibility should not be hastily discarded. Indeed, it is  
only recently that a couple of well-known globular clusters (Pal 5,  
Odenkirchen et al. 2001, and NGC5466, Belokurov et al. 2006a) have  
been convincingly shown to exhibit the telltale signatures of tidal  
stripping.  The unambiguous identification of tidally-induced features  
is even harder when the orbits are eccentric, since in that case the  
episodes of stripping are short and restricted to pericentric  
passages.  A dwarf is expected to relax back to equilibrium on a  
dynamical timescale, erasing quickly any signs of the interaction from  
the main body of the dwarf, where crossing times are shortest.  
  
This difficulty has not deterred a rich body of work from postulating  
links between tides and the present-day structure of Galactic  
dSphs. This is typically argued on the basis of the projected stellar  
density profile, $\Sigma(R)$, where deviations or ``breaks'' from a  
smooth profile are often interpreted as evidence for the effect of  
tides.  
  
Such interpretation is encouraged by the success of King models (King  
1966) at reproducing the surface brightness profile of most dSphs. In  
particular, King models describe well the central region of nearly  
constant density (the ``core'') as well as the precipitous decline in  
the outer regions usually ascribed to a ``tidal radius'' imposed by  
the gravitational field of the Galaxy. ``Features'' in the outer profile,  
such as an upturn in $\Sigma(R)$ relative to the best-fitting King  
model, are therefore interpreted as caused by ``unbound'' stars  
leaving the dwarf.  
  
The latter assertion, however, is quite difficult to distinguish from  
more prosaic interpretations where the observed complexities in  
$\Sigma(R)$ are ascribed to radially-dependent variations in star  
formation history: in other words, a dSph may deviate from a King  
model simply because it was born like that.  The recent discovery of  
distinct (bound) dynamical populations in some Local Group dSphs  
(Sculptor, Tolstoy et al. 2004; Canes Venatici I, Ibata et al. 2006;  
Fornax, Batagglia et al. 2007; Andromeda II, McConnachie, Arimoto \&  
Irwin 2007), and the breaks in $\Sigma(R)$ that such populations can  
produce (see, e.g., Figure~2 of McConnachie, Pe\~narrubia \& Navarro  
2007) suggest that this alternative explanation should be taken into  
consideration.  
   
Although it might be difficult to determine conclusively whether the  
properties of a dSph have been modified by the action of tidal forces,  
it is clear that at least in some cases, like Sagittarius, tides are  
at work and will determine the dwarf's ultimate structure. This brings  
about interesting questions. One relates to the link between tides and  
the ``King-like'' $\Sigma(R)$ profile of many dSphs. How resilient is  
$\Sigma(R)$ to the effect of tides?  Would a dSph where stars  
initially follow a King-like model evolve away from it, approaching,  
for example, a power-law profile as it soaks gravitational energy from  
tides? Can this be used to gauge the dynamical importance of tides in  
the evolution of a dSph?  
   
We explore these issues here using cosmologically-motivated models for  
dSphs. These models are similar to those introduced by Pe\~narrubia,  
McConnachie \& Navarro (2007, hereafter PMN, see also Strigari et al  
2007), who used them to constrain the dark matter halos of Local Group  
dSphs under the assumption that tides have not perturbed drastically  
the halo properties. This approximation, however, is unlikely to hold  
for many dSphs and therefore it is imperative to examine the effects  
of tides on such estimates.  
  
Following PMN, we assume that dSphs are dark matter-dominated systems  
where stars are ``King-model''-distributed tracers embedded within a  
cold dark matter (CDM) halo. These systems are on eccentric orbits  
that bring them periodically close enough to the center of the  
Galactic potential for tides to act. We explore a wide range of  
orbital parameters, as well as the dependence of our results on the  
degree of segregation of the stellar component within the dark  
halo. Our analysis focuses on the equilibrium structure of the  
remnants, and are thus best applied to systems near the apocenter of  
their orbits and away from regions where tides are strongest. We will  
address in a separate contribution (Pe\~narrubia et al., {\it in  
prep.}) the issue of transient departures from equilibrium and their  
use as diagnostics of the importance of tides.

One limitation is that our dSph models are evolved individually  
within a rigid Galactic potential, and therefore our computations  
neglect the effects of dynamical friction (e.g. Pe\~narrubia et al. 2002, 2004, 2006) and of interactions between  
satellites  (Pe\~narrubia \& Benson 2005). Since we concentrate on some of the faintest (and thus  
presumably least massive) systems known, these omissions should not  
prove deleterious to our conclusions.  
   
This paper is organized as follows. In Sec.~\ref{sec:model} we  
introduce our dSph modeling and describe in detail the numerical  
technique.  Sec.~\ref{sec:tidev} presents our main results regarding  
the dynamical evolution of dSphs driven by tides. In  
Sec.~\ref{sec:lgdsph} we apply our results to the population of Local  
Group dSphs and newly-discovered ultra-faint galaxies. We end with a  
brief summary in Sec.~\ref{sec:sum}.  
   
\section{Numerical modeling}   
\label{sec:model}   
   
Our model assumes that the stellar component of dSphs follow a King  
(1966) model embedded within a CDM halo. The dark halo is modeled as  
an N-body realization of a spherical Navarro, Frenk \& White (1996,  
1997) (hereafter NFW) profile. Stars are assumed to   
contribute negligibly to the potential, and may therefore be followed  
by assigning an energy-dependent ``mass-to-light ratio'' to each dark  
matter particle in the equilibrium halo. The dwarf is assumed to orbit  
within a massive host system which we model, for simplicity, also as a  
rigid NFW potential.   
  
Our approach is thus similar to that of Hayashi et al (2003), Bullock  
\& Johnston (2005), and Kazantzidis et al (2004), but we  
concentrate the analysis on the evolution of the ``stellar  
component'' of the main body of the dwarf. Further numerical details  
are given in the remainder of this section.  
   
\begin{figure}   
\plotone{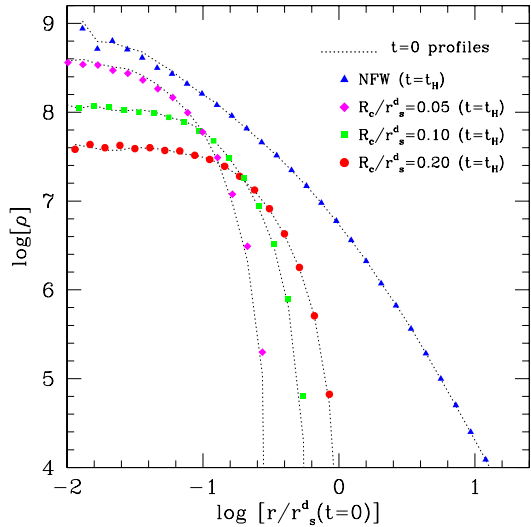}   
\caption{Density profiles of the King-NFW dwarf galaxy model   
considered in this paper (arbitrary units). The dark halo is described as an NFW model   
(top curve); the stellar component is assumed to follow a King model   
constructing by sampling the NFW model in energy space (other   
lines). Three different choices for the spatial segregation between   
stars and dark matter are shown: $R_c/r_s=0.05$, $0.1$, and   
$0.2$. Dotted lines correspond to the models as initialized by the   
N-body procedure. Symbols show the configuration of the same models,   
but after evolving the system for 14 Gyr in isolation. The good   
agreement between lines and symbols indicate that the dSph model is   
in equilibrium and free from numerical artifact induced by numerical   
limitations down to scales comparable to the grid size of the highest   
resolution zone used in {\sc Superbox}, $r_s/126$.}   
\label{fig:dens_is}   
\end{figure}   
\begin{figure*}   
\plotone{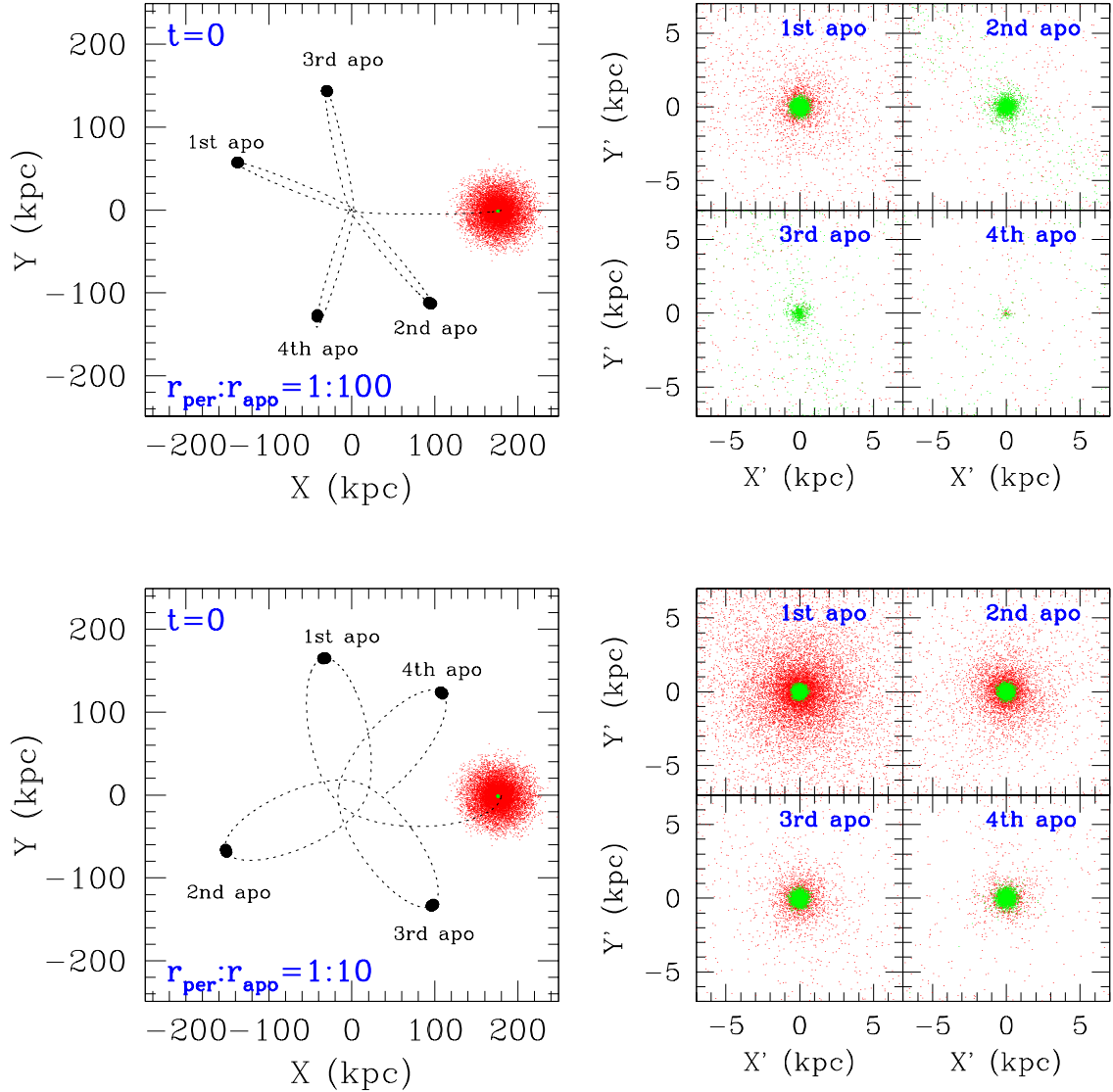}   
\caption{Projection onto the orbital plane of two dSph models with  
different orbital pericenter. Left panels show the initial systems at  
scale.  Right-hand panels show the inner 15 kpc of the dwarf at  
various times corresponding to four consecutive apocenters for each  
orbit. Red and green dots denote, respectively, dark and stellar  
particles. Note how the stripping proceeds from the outside in and  
that a large fraction of the dark matter halo must be tidally stripped  
before stellar mass loss begins.}  
\label{fig:xy}   
\end{figure*}   
   
\begin{figure*}   
\plotone{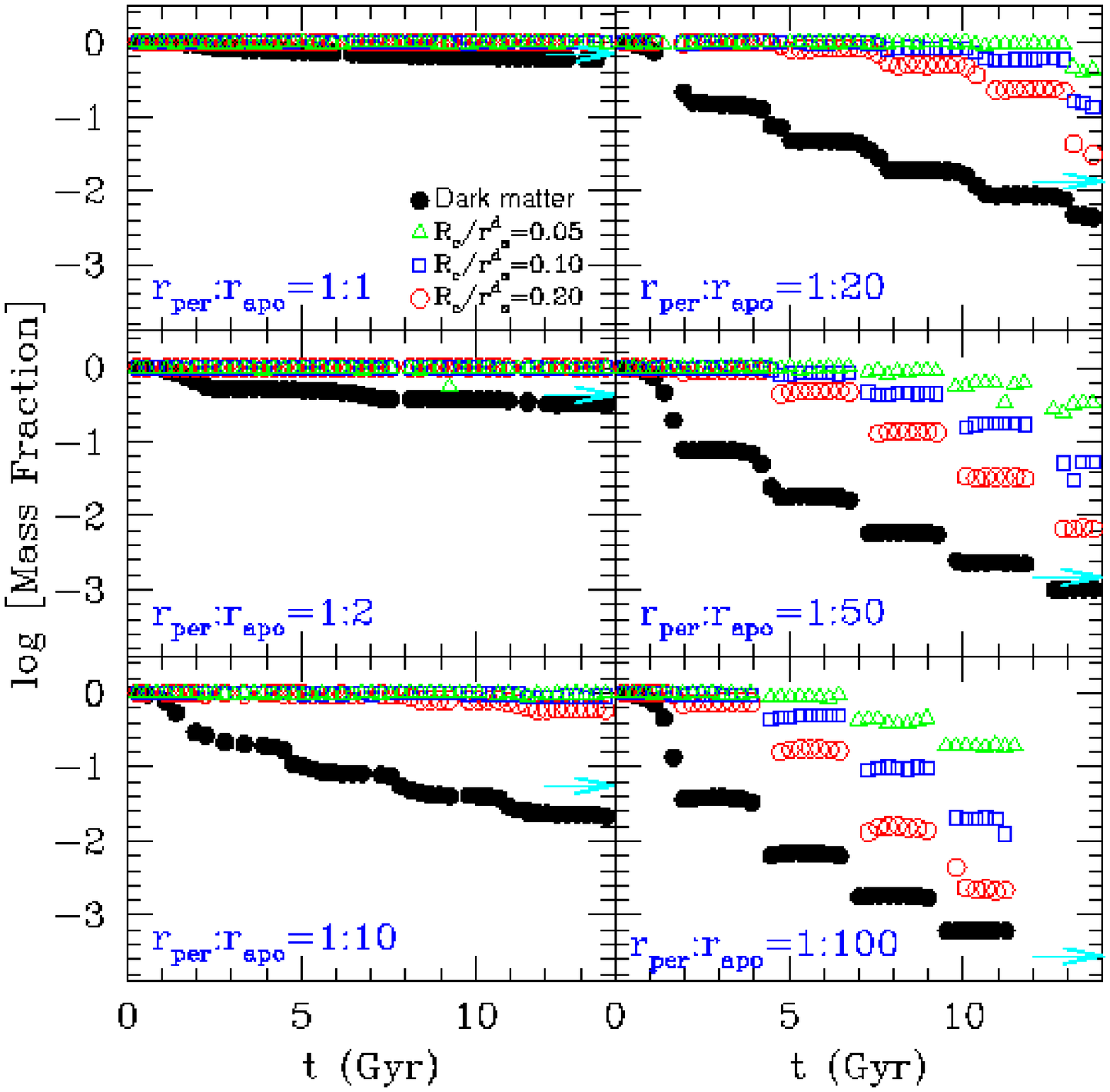}   
\caption{Bound mass fraction as a function of time for different  
orbits and stellar segregations. Arrows indicate the mass fraction  
within the theoretical ``tidal radius'' estimated at pericenter (see  
text for details). Note that a large fraction of the halo mass must be  
tidally stripped ($\sim 90\%$) before stars begin to be  
stripped. However, when stars are stripped, they are lost more  
efficiently than the remainder of the dark halo. See text for  
details. }   
\label{fig:mloss}   
\end{figure*}   
\subsection{The host galaxy potential}  
\label{ssec:hostmodel}  
  
The host galaxy potential follows from the NFW density profile, which  
may be expressed as,  
\begin{eqnarray}  
\rho_{\rm NFW}(r)=\frac{M_{\rm vir}}{4\pi r_s^3} \ \frac{(r/r_s)^{-1}  
(1+r/r_s)^{-2}} {\ln(1+c)-c/(1+c)},  
\label{eq:rhoh}  
\end{eqnarray}  
where $M_{\rm vir}$ is the mass within the virial radius, $r_{\rm  
vir}$, $r_s$ is a scale radius and $c$ is the NFW concentration  
parameter, defined as $c\equiv r_{\rm vir}/{r_s}$.  
  
The virial radius is defined so that the mean over-density relative to  
the critical density is $\Delta$,  
\begin{equation}  
{M_{\rm vir} \over (4/3) \pi r_{\rm vir}^3}= \Delta \ \rho_{\rm crit},  
\end{equation}  
where $\rho_{\rm crit}=3 H_0^2/8 \pi G$, and $H_0=100\, h$ km/s/Mpc is  
the present day value of Hubble's constant.  
  
The choice of $\Delta$ varies in the literature, with some authors  
using a fixed value, such as NFW, who adopted $\Delta=200$, and others  
who choose a value motivated by the spherical collapse model, where  
(for a flat universe) $\Delta\sim 178\, \Omega_{\rm m}^{0.45}$ (Lahav  
et al 1991, Eke et al 1996). The latter gives $\Delta=95.4$ at $z=0$  
in the concordance $\Lambda$CDM cosmogony, which adopts the following  
cosmological parameters: $\Omega_{\rm m}=0.3$, $\Omega_\Lambda=0.7$,  
$h=0.7$, consistent with constraints from CMB measurements and galaxy  
clustering (see, e.g., Spergel et al 2006 and references therein).  
  
Our models are purely gravitational and therefore scale-free. When  
convenient for interpretive purposes, we shall scale our numerical  
units to physical parameters assuming that the host galaxy potential  
has $M_{\rm vir}=7.55\times 10^{11} M_\odot$, $r_{\rm vir}=233.0$  
kpc and $r_s=8.18$ kpc, assuming $\Delta=101.1$. This  
corresponds to a virial velocity of $V_{\rm vir}=117.8$ km/s and a  
circular velocity profile that reaches its peak, $V_{\rm max}=188.1$  
km/s, at $r_{\rm max} \approx 2\, r_s = 16.4$ kpc. An NFW profile is  
fully characterized by a pair of independent parameters, for example,  
$M_{\rm vir}$ and $c$, or $r_{\rm max}$ and $V_{\rm max}$.  
  
The potential corresponding to an NFW halo is given by   
\begin{equation}  
\Phi_{\rm NFW}(r)=-\frac{M_{\rm vir}}{r} \ \frac{\ln(1+r/r_s)}{\ln(1+c)-c/(1+c)},  
\label{eq:phi}  
\end{equation}  
and is assumed to remain constant during the evolution of the  
dwarf.  
  
For simplicity, our host galaxy model does not include the 
presence of a disk or bulge. The presence of these components would 
break the spherical symmetry of the host halo and increase the number 
of free parameters needed to describe it in detail. However, as we 
shall show below, the effects of tides on the structure of dSphs seem 
to depend solely on the total amount of mass lost to tides, and not on 
the details of how it was stripped. In this sense, our results are 
likely to depend largely on the strength of the tides and only weakly 
on the exact potential shape of the host galaxy, and our assumption of 
a spherically symmetric halo should still capture the most relevant 
features of this process. 
  
\begin{figure}  
\plotone{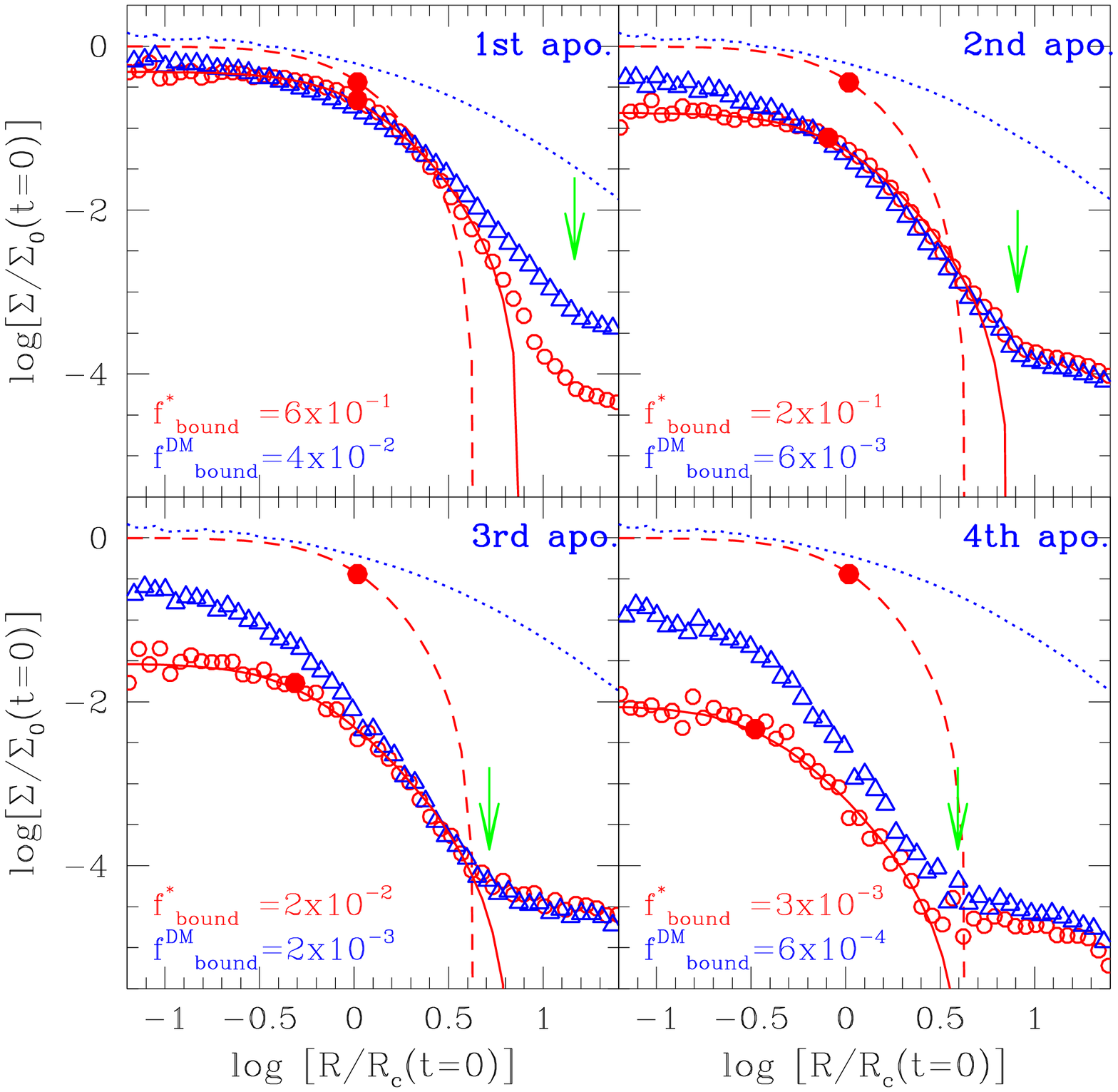}  
\caption{Surface density profiles of our dSph model (assuming a  
stellar segregation of $R_c/r_s^d=0.20$) moving on a highly eccentric  
orbit ($r_{\rm per}$:$r_{\rm apo}=1$:$100$) at different orbital  
apocenters. Stellar and dark matter profiles are denoted with open  
circles and triangles, respectively. For comparison we also plot the  
initial profiles in each panel (dotted lines). At each snapshot we fit  
a King profile to the stellar component (solid lines). Solid dots show  
the resulting value of the core radius ($R_c$). Arrows indicate the  
position of $R_{\rm eq}$, where the crossing time equals the time  
elapsed since pericenter. Labels in each panel indicate the remaining  
fraction of bound stellar and dark mass, respectively. See text for  
details.}  
\label{fig:dens_evol}  
\end{figure}  
  
\subsection{The dSph N-body realization}  
\label{ssec:dsphmodel}  
  
Our dwarf galaxy N-body model is constructed by generating an  
equilibrium $5$ million-particle realization of the NFW profile, using  
a code kindly made available by S.Kazantzidis (Kazantzidis et  
al. 2004, 2006). Since the NFW profile has infinite radial extent and  
divergent total mass, the profile is truncated outside the virial radius.  
  
We have chosen parameters for the dwarf consistent with those expected  
for dSph halos (see, e.g., PMN). Using the same physical scaling  
adopted for the host galaxy (\S~\ref{ssec:hostmodel}) we adopt $M_{\rm  
vir}=3.7\times 10^9 M_\odot$, $r_{\rm vir}=39.6$ kpc, and $c=20$. This  
implies $V_{\rm vir}=20.0$ km/s; $r_s=2.0$ kpc; $V_{\rm max}=28.6$ km/s;  
and $r_{\rm max}=4.0$ kpc.  
  
We emphasize, however, that our models are scale free, so that the  
only relevant parameters in the calculations are the {\it relative}  
masses and sizes of the host and dwarf, which may be expressed by the  
ratio of their maximum circular velocities, $V_{\rm max}^{\rm  
host}/V_{\rm max}^{\rm dSph}=6.6$ and by $r_{\rm max}^{\rm host}/r_{\rm  
max}^{\rm dSph}=4.1$. In what follows, a ``d'' or ``h'' upper script  
will be used to denote quantities corresponding to the ``dwarf'' or  
the ``host'' models, respectively.  
  
Our model assumes that stars are mass-less tracers of the potential,  
and therefore we track the stellar component using a subset of dark  
matter particles, chosen so that they describe a King model in  
equilibrium within the dwarf halo. We follow the method outlined by  
Bullock \& Johnston~(2005), and assign weights to dark matter  
particles so as to select $f_K(\epsilon)/f_{\rm NFW}(\epsilon)\times  
N(\epsilon)$ stars from each bin in relative energy, $\epsilon\pm  
\delta \epsilon$.  Here $f_K$ and $f_{\rm NFW}$ are the isotropic  
distribution functions for the stellar and the dark matter particles  
calculated according to  
\begin{eqnarray}  
f_i(\epsilon)=\frac{1}{8\pi^2}\ \bigg[\int_0^\epsilon  
\frac{d^2\rho_i}{d\Psi^2}\frac{d\psi}{\sqrt{\epsilon-\Psi}}+\frac{1}{\sqrt{\epsilon}}\bigg(\frac{d\rho_i}{d\Psi}\bigg)_{\Psi=0}\bigg],  
\label{eq:edding}  
\end{eqnarray}  
where the subscript $i$ denotes that $\rho$ can be either an NFW or a  
King profile, and $\Psi$ is the total gravitational potential.  
  
The result is a set of $N_\star$ dark matter particles that follow a  
King (1962) density distribution,  
\begin{eqnarray}  
\rho_\star=\frac{K}{x^2}\  
\bigg[\frac{\cos^{-1}(x)}{x}-\sqrt{1-x^2}\bigg],~~x\equiv\bigg[\frac{1+(r/r_K)^2}{1+(r/r_t)^2}\bigg]^{1/2},  
\label{eq:rhok}  
\end{eqnarray}  
where $r_K$ and $r_t$ are the King and tidal radii and $K$ is an  
arbitrary constant. For all our dSph models we have fixed $r_t/r_K=5$.  
  
To compare with observations it is easier to use the projected King  
(1962) surface density  
\begin{eqnarray}   
\Sigma (R)= k \Bigg\{  \frac{1}{ \big[1+(R/R_c)^2\big]^{1/2} }-\frac{1}{ \big[1+(R_t/R_c)^2\big]^{1/2} }  \Bigg\}^2  
\label{eq:Sigma}   
\end{eqnarray}   
where $R$ is the projected radius and $k$ an arbitrary constant. For  
our choice of $r_t/r_K=5$, we have that our King models have  
``concentration'' $c_K\equiv R_t/R_c=5$, and central surface  
brightness given by $\Sigma_0=k\big(1- 1/\sqrt{1+c_K^2}\big)^2$.

With these choices, the whole procedure depends solely on the degree  
of spatial segregation between stars and dark matter in the dwarf,  
which we may express as the ratio between the stars' core radius and  
the halo NFW scale radius, $R_c/r_{s}^{\rm d}$. As discussed by PMN,  
this segregation parameter may be constrained by using the cosmological relationships between the mass and concentration of NFW halos obtained from N-body simulations (e.g. NFW, Eke,  
Navarro \& Steinmetz 2001, Bullock et al 2001).  
PMN find values of $R_c/r_s^d$ in the range  
$0.05$--$0.20$ (or, equivalently, $R_c/r_{\rm max}^d\simeq  
0.025$--$0.1$), which implies that the stellar King  
models are deeply embedded within their NFW halos. This results in   
velocity dispersion profiles, $\sigma_p(R)$, that, in concordance with observations, are nearly  
flat almost out to the ``tidal'' radius.

The more deeply segregated the stellar component is, the fewer  
particles are available to trace it, and therefore large numbers of  
particles are required in order to resolve the innermost regions of  
halos where the stars are thought to reside.  Using a total number of  
$5\times 10^6$ dark matter particles for the NFW profile we obtain  
stellar components traced by $1.95\times 10^4$, $6.86\times 10^4$, and  
$1.88\times 10^5$ particles, for $R_c/r_s=0.05$, $0.10$, and $0.2$,  
respectively.  
  
\subsection{The N-body code}  
\label{ssec:numsetup}  
  
We follow the evolution of the dSph N-body model in the host potential  
using {\sc Superbox}, a highly efficient particle-mesh gravity code  
(see Fellhauer et al. 2000 for details). {\sc Superbox} uses a  
combination of different spatial grids in order to enhance the  
numerical resolution of the calculation in the regions of interest.  
In our case, {\sc Superbox} uses three nested grid zones centered on  
the highest-density particle cell of the dwarf.  This center is  
updated at every time step, so that all grids follow the satellite  
galaxy along its orbit.  
  
Each grid has $128^3$ cubic cells: (i) the inner grid has a  
spacing of $dx=r_s^d / 126\simeq 8\times 10^{-3} \, r_s^d$ and is meant to  
resolve the innermost region of the dwarf. As discussed above, we  
assume that stars are heavily segregated relative to the dark matter,  
so that the inner grid spacing is of order $dx \simeq  
R_c/6$--$R_c/25$.  (ii) The middle grid extends to cover the whole  
dwarf, with spacing $r_{\rm vir}^d/ 126$. (iii) The outermost grid  
extends out to $50\times r_{\rm vir}^d$ and is meant to follow particles  
that are stripped from the dwarf and that orbit within the main  
galaxy..  
  
{\sc Superbox} uses a leap-frog scheme with a constant time-step to  
integrate the equations of motion for each particle. We select the  
time-step according to the criterion of Power et al (2003); applied to  
our dwarf galaxy models, this yields $\Delta t=4.6$ Myr for the  
physical unit scaling adopted in Sec.~\ref{ssec:hostmodel}.  
  
\subsection{Tests of the dSph model}  
\label{ssec:tests}  
  
We have checked explicitly that our procedure for building the dSph  
model leads to a system that is in equilibrium and that does not  
evolve in isolation away from the prescribed configuration. This is shown  
in Fig.~\ref{fig:dens_is}, where we show the density profile of the  
NFW profile (top dotted line), as well as the profile of the  
King-model subsets, for three choices of $R_c/r_s$ (other lines,  
arbitrary vertical normalization).  The symbols show the same  
profiles, but after evolving the system in isolation for 14 Gyr with  
{\sc Superbox}. The good agreement between lines and symbols shows  
that our numerical choices are appropriate, and that the evolution of  
the N-body model, on the scales of interest, is free from artifact  
induced by the finite spatial and time resolution of the calculation.  
  
\section{Tidal Evolution of dwarf spheroidals}  
\label{sec:tidev}  
  
\subsection{The Orbits}  
\label {ssec:orbits}  
  
In order to explore the tidal evolution of dSphs we have placed the  
dwarf galaxy model in orbit within the potential of the host galaxy  
assuming 6 different combinations of apocentric/pericentric ratio. All  
6 orbits have apocenters  $r_{\rm apo}=0.77 r_{\rm vir}^{\rm h}=180$ kpc.  The strength of the tidal  
forcing is thus controlled by the pericentric radius, which we vary  
from $r_{\rm per}=r_{\rm apo}$ (a circular orbit) to $1/2$,  
$1/10$, $1/20$, $1/50$, and $1/100$ of $r_{\rm apo}$. Scaling  
this to our chosen physical units, the orbit with the smallest  
pericenter reaches within $\sim 2$ kpc of the center of the host, but  
one should bear in mind that the scaling is arbitrary, and that it is  
best to regard the series of simulations as one of increasing tidal  
strength for eccentric orbits of similar orbital period.  
  
We may quantify the expected strength of the tides by computing the  
theoretical ``tidal radius'' of the dwarf galaxy model if it were  
placed on a circular orbit at pericenter. Defining $r_{\rm tid}$ as  
$\langle \rho^{\rm d} \rangle (r_{\rm tid})=3\, \langle \rho^{\rm  
h}\rangle (r_{\rm per})$ (where $\langle \rho \rangle(r)$ denotes mean  
enclosed density within $r$) we may estimate the total amount of mass,  
$M_{\rm tid}\equiv M^d(r_{\rm tid})$, that the satellite might be  
expected to retain after completing several orbits in the host  
potential.  
  
The results are compiled in Table~\ref{tab:mod}. Except for the  
circular orbit case, it is clear from inspecting these values that we  
expect tides to have a strong influence on the evolution of the  
dwarf. For example, more than half of the initial mass is expected to  
be lost in the $r_{\rm per}$:$r_{\rm apo}=1$:$2$ orbit, growing to  
$99.97\%$ in the case of the most extreme orbit ($1$:$100$). In all  
cases we follow the evolution of the satellite for about $t_H=14$ Gyr (in  
our scaled units), which corresponds to roughly $3$--$5$ radial  
orbital periods.  
  
We show for illustration the orbital paths of two simulations in the  
left-hand panels of Fig.~\ref{fig:xy}. These correspond to the $r_{\rm  
per}$:$r_{\rm apo}=1$:$100$ and $1$:$10$ cases, respectively. The dark  
matter particles of the dSph model are shown in red, and are at  
scale. The stellar component is shown in green (assuming $R_c/r_s^{\rm  
d}=0.2$), and is barely seen as a dot at the center of the dwarf; this  
gives a visually intuitive impression of how deeply segregated the  
stars are relative to the dark matter in our models.

\begin{figure}  
\plotone{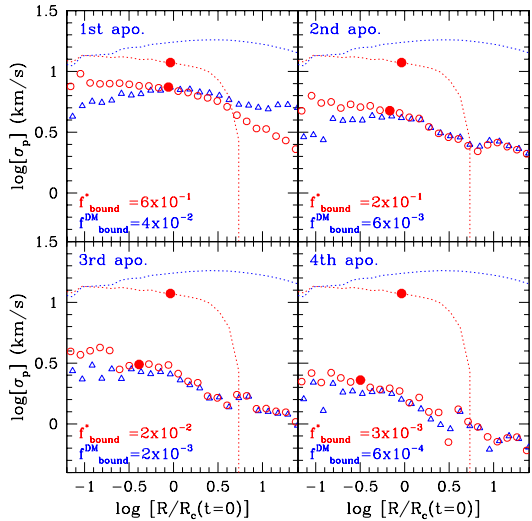}  
\caption{As Fig.~\ref{fig:dens_evol}, but for the projected velocity  
dispersion profiles. Only bound particles are considered in these  
panels. Note that tidal stripping leads to a monotonic decrease of  
$\sigma_p(R)$ at all radii. However, even after extreme mass loss, the  
shape of the velocity dispersion profile of the bound stellar remnants  
remains approximately constant out to roughly $\sim 3\, R_c$.  }  
\label{fig:s_evol}  
\end{figure}  

\subsection{Mass loss}  
\label{ssec:mloss}  
  
As expected, tides are strongest at pericenter and trigger recurring  
episodes of mass loss at every pericentric passage. (The obvious  
exception is the circular orbit case, where little stellar mass loss is  
expected, see Table~\ref{tab:mod}.)  The progressive stripping of the  
satellite is shown in Fig.~\ref{fig:mloss}, where the solid circles  
show the fraction of mass that remains bound to the satellite as a  
function of time. The pericentric passages are easily recognizable in  
this plot as the times when sudden drops in bound mass occur. Except  
at pericenter, the bound mass of a satellite remains fairly constant  
during the orbit, suggesting, as indicated above, that tides act  
nearly impulsively along the chosen orbits and have little influence  
on the main body of the dSph at times other than pericenter.  
  
The inner regions of the halo are remarkably resilient to disruption  
by such ``tidal shocks''. Indeed, after 14 Gyr all of our dSph models,  
except for the most extreme orbit, still retain a fraction of their  
mass. The retained mass fraction after completing $3$-$5$ orbits is in  
reasonable agreement with the initial mass within the ``tidal  
radius'', $r_{\rm tid}$, which is shown by the rightward pointing  
arrows in Fig.~\ref{fig:mloss}. Despite this coincidence, there is no  
sign that the retained mass is converging to a well defined value in  
the case of the three most eccentric orbits.  The dSph halo is fully  
disrupted during the 5th pericentric passage of the $1$:$100$ orbit  
although one would have expected that $\sim 0.03\%$ (about 2000  
particles) should remain bound in this case. A similar result is  
obtained for the $1$:$50$ orbit when extending the simulation beyond  
the nominal $14$ Gyr timespan. This suggests that the ``tidal mass''  
listed in Table~\ref{tab:mod} actually underestimates the mass that  
remains attached to a dSph after repeated pericentric passages.  
  
Mass loss loss affects predominantly the outer regions of the dSph  
halo. This is shown by the open symbols in Fig.~\ref{fig:mloss}, which  
indicate the evolution of the {\it stellar} bound mass fraction, for  
three different values of the segregation parameter; $R_c/r_s^d=0.05$,  
$0.1$, and $0.2$. The stellar mass loss, as a whole, is always less  
pronounced than the dark matter's, as expected for a stellar system  
that is deeply embedded within its dark halo. Depending on how  
segregated the stars are, a dSph may lose $\sim 90\%$ of its original  
dark mass and still retain all of its stars (see, e.g., the bottom  
left panel in Fig.~\ref{fig:mloss}). For the range of segregations  
adopted here, a dSph must lose at least $90\%$ of its mass before  
starting to shed its stars.  
  
The outside-in ``onion-peel'' stripping of the dSph may be appreciated  
in the right-hand panels of Fig.~\ref{fig:xy}, where we show, at  
consecutive apocenters of the orbit, the distribution of dark matter  
particles (in red) and stars (in green) in a $15$ kpc box centered on  
the dwarf. This shows clearly how a halo is gradually stripped of its  
mass from the outside in. The bottom panels of this figure, in  
particular, illustrate how a dSph may be stripped of much of its halo  
whilst leaving most of its stars bound. The stellar components  
of dwarf halos are clearly extremely resilient to tidal stripping.  
  
Finally, another interesting result may be gleaned from  
Fig.~\ref{fig:mloss} by noting that, when stars are lost, the  
``jumps'' in bound stellar mass are larger than those associated with  
the dark matter as a whole. This implies that, when the stellar  
component actually begins to disrupt, it does so more rapidly than the  
remainder of its dark matter halo. This has interesting consequences  
for the evolution and interpretation of the dSph structural  
parameters, an issue to which we return below.  
  
\subsection{Evolution of the density profile}  
\label{sec:evdprof}  
  
Fig.~\ref{fig:dens_evol} shows the evolution of the projected density  
profile of dark matter (triangles) and stars (circles) for the most  
extreme orbit ($r_{\rm per}$:$r_{\rm apo}=1$:$100$) probed in our  
series. The stellar profile is shown for the case when the segregation  
parameter is $R_c/r_s^d=0.2$. The system is shown at four  
consecutive apocentric passages, when the main body of the dSph has  
had some time to relax after being acted on by tides at pericenter. The  
dotted lines are the same in each panel, and show the initial dark  
matter and stellar profiles, for ease of reference. The vertical  
normalization of the initial stellar profile is arbitrary, but is the  
same in all panels.  
  
This figure illustrates a few interesting points. One is that it is  
possible to identify two distinct regions on the basis of the shape of  
the density profile: an inner region where the profile varies smoothly  
with radius, and an outer region where an ``excess'' of mass is  
present relative to a naive extrapolation of the inner profile. The  
transition radius coincides roughly with the arrow marking the  
distance from the center where the crossing time equals the time  
elapsed since pericenter, $t-t_{\rm per}=t_{\rm cross}(R_{\rm eq})=  
R_{\rm eq}/\sigma_0$, where $\sigma_0 \equiv \sigma_p(R=0)$ is the  
central line-of-sight stellar velocity dispersion.  
  
Particles inside $R_{\rm eq}$ are thus within the region where enough  
time has elapsed for equilibrium to be re-established, whereas the  
outer region still contains a large number of weakly-bound particles  
that are still moving out and have yet to settle into their new orbits  
(Aguilar \& White 1985, 1986, Navarro 1990). These particles are  
responsible for the transient ``excess'' of mass in the outer  
regions. The location of this excess moves gradually outward, and it  
would therefore be difficult to detect unless a dwarf is inspected  
soon after pericenter. Note that stars in the outer envelope are not  
{\it necessarily} unbound, and that the presence of this ``tidal  
break'', as the outer excess is sometimes referred to, does not  
necessarily imply the presence of an unbound population.  
  
A second point of note is that, within $R_{\rm eq}$, the stellar  
density profile is very well approximated by a King model, as shown by  
the solid line fits to the inner open circles in  
Fig.~\ref{fig:dens_evol}. The solid dot indicates the core radius of  
the fit; intriguingly, this seems to evolve rather weakly, changing by  
less than a factor of $\sim 3$ even when the stellar component has  
been stripped of roughly $99.7\%$ of its original mass. Stellar mass  
loss thus correlates almost inversely proportional to the central  
surface density, which drops by about two decades after completing 4  
orbits.  
  
The change in projected density is much less pronounced in the dark  
matter, emphasizing that, when stars are lost to tides, they do so  
with higher efficiency than the remaining dark matter. This is because  
the dark matter profile is ``cuspy'' whereas that of the stars is  
``cored'': the dark matter in the very inner regions is therefore more  
resilient to tides than stars. As a result, {\it tides lead generally  
to an increase in the mass-to-light ratio} of a dSph remnant. This  
statement depends critically on the cuspy inner profile of the dark  
matter, and may therefore be regarded as a genuine prediction of the  
$\Lambda$CDM scenario.  
  
\subsection{Evolution of the velocity dispersion profile}  
\label{sec:evsigprof}  
  
The evolution of the velocity dispersion profile of stars and dark 
matter is shown in Fig.~\ref{fig:s_evol}. As in 
Fig.~\ref{fig:dens_evol}, the dotted lines indicate the initial 
profiles and are the same in all panels. The solid dot marks the core 
radius of the best King model fits to the equilibrium region, i.e., 
$R<R_{\rm eq}$. Unlike the plots in Fig~\ref{fig:dens_evol}, we use 
only {\it bound} particles to compute the velocity dispersion.   
As discussed in the literature, the presence of unbound stars would 
increase the velocity dispersion estimates and may lead to 
``features'' in the velocity dispersion profile. Observationally, the 
interpretation of these features is difficult, since such features may 
be produced by stars stripped from the dwarf or by unrelated objects 
projected onto the line of sight (e.g. Johnston et al. 2002, Read et 
al. 2006, Fellhauer \& Kroupa 2006, Klimentowski et 
al. 2007). However, considering only bound particles, the velocity 
dispersion profiles show no obvious feature separating the equilibrium 
region from the unrelaxed envelope. Thus, the profiles shown in 
Fig.~\ref{fig:dens_evol} would be expected in galaxies that have had 
time to relax after the pericentric passage and whose unbound 
component has already evacuated the progenitor system. 
 
For Figures~\ref{fig:dens_evol} and~\ref{fig:s_evol} we have selected a relatively extended stellar component ($R_c/r_s^d=0.2$) to show the enhanced effects of tidal interactions. This results in an initial velocity dispersion profile that is slightly declining at large radii. Measurements of $\sigma_p(R)$ in the Local Group dSphs show nearly flat velocity dispersion profiles within at least $3\, R_c$ (see \S~1). As discussed by PMN, that condition implies $R_c/r_s^d\lesssim 0.1$ for our initial models.

Remarkably, Fig.~\ref{fig:s_evol} shows that tidal mass stripping does not alter the velocity dispersion profile of the bound stellar component, even after the dwarf galaxy loses 99\% of the original stars: at $R\sim 3\, R_c$ drops by less than $40\%$ from its central value in all four panels of this Figure.
The  
central velocity dispersion, on the other hand, drops by a factor of  
$\sim 5$ after the system has lost $99.7\%$ of its stars, a result that  
will prove useful when trying to interpret observationally the tidal  
remnants of Local Group dSphs. 
  
Note that tides preserve the ``King-like'' structure of the stellar  
component and that, therefore, the bound remnants of stripped dwarf  
galaxies will show no obvious signature of tidal effects once  
they have been able to relax to virial equilibrium after the latest  
pericentric passage.  
  
\subsection{Evolution of stellar structural parameters}  
\label{sec:evststructpar}  
  
As shown in Fig.~\ref{fig:dens_evol}, NFW-embedded King models are  
very resilient to the effect of tides; even after being stripped of  
99\% of its stars, the remaining bound stellar component resettles  
into a King-like configuration in just one dynamical time. The change  
in the best fitting King-model parameters will depend on the strength  
of the tidal interaction as well as on the spatial segregation between  
stars and dark matter. The more segregated the stellar component the  
harder it is for tides to affect it.  
  
\begin{figure*}  
\plotone{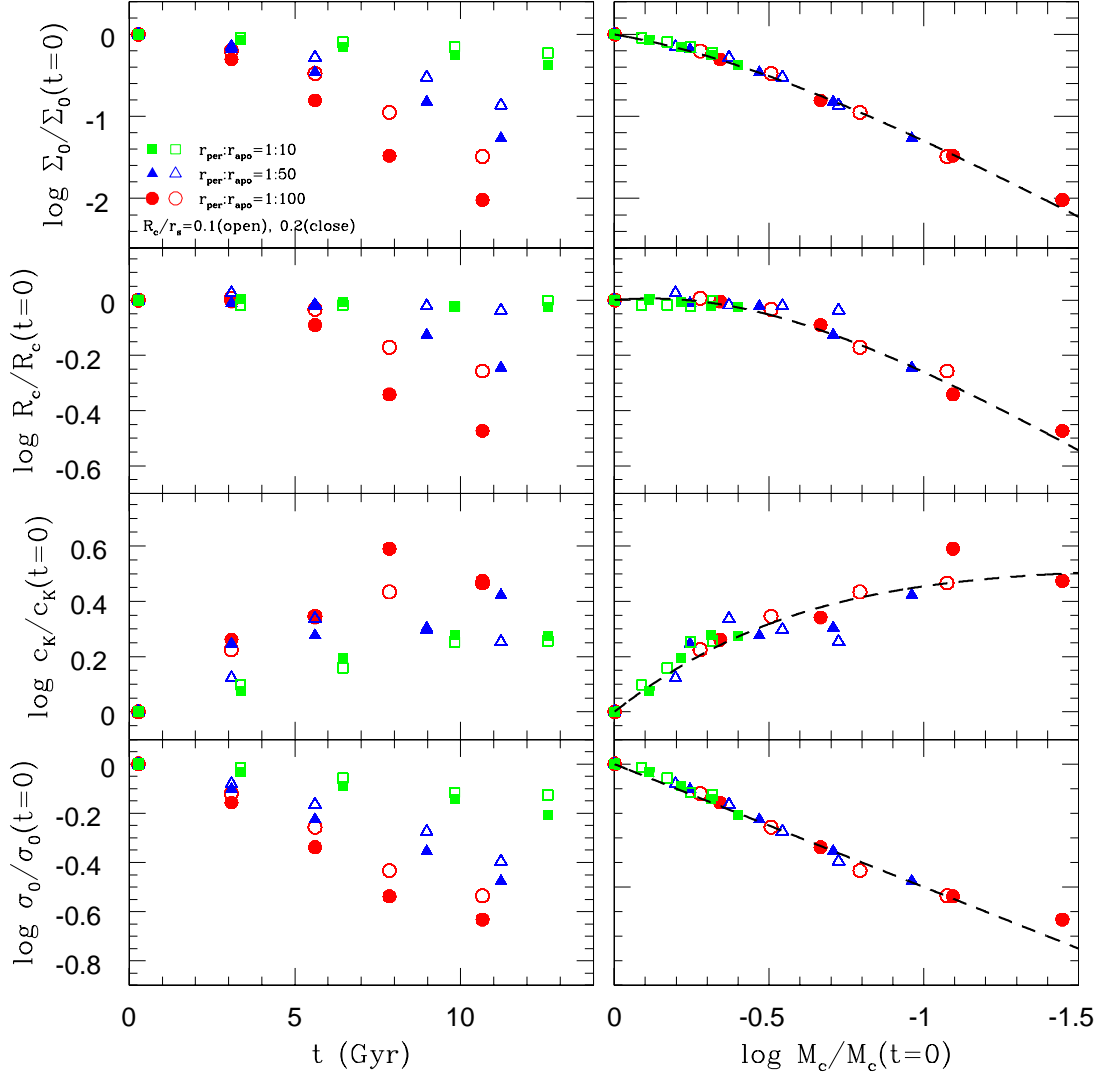}  
\caption{Evolution of the parameters of the King models that best fit  
the inner ($R<R_{\rm eq}$) stellar surface density profile:  
$\Sigma_0$, $R_c$, and $c_K=R_t/R_c$. Also shown is the central  
projected velocity dispersion, $\sigma_0$. These parameters are shown  
as a function of time (panels on the left) and as a function of the  
total bound mass within the initial core radius, $M_c/M_c(t=0)$  
(panels on the right). Different symbols correspond to different  
orbits and stellar segregations measured at consecutive  
apocenters. Dashed lines in the right-hand panels show our empirical  
fits (see eq.~\ref{eq:gx} and Table~\ref{tab:fit}). Note that the  
evolution of the stellar parameters only depends on the total amount  
of mass lost from within the luminous radius and not on the details of  
how this mass was stripped.}  
\label{fig:magic}  
\end{figure*}

This is demonstrated in the panels on the left of  
Figure~\ref{fig:magic}, which show, as a function of time, the  
evolution of the central surface density, $\Sigma_0$, core radius,  
$R_c$, and King ``concentration'', $c_{\rm k}=R_t/R_c$, for three of  
the six orbits surveyed in our study. Different symbols correspond to  
different orbits, and are shown at consecutive orbital apocenters,  
when the main body of the dSph is close to equilibrium. Open and  
filled symbols correspond to assuming a radial segregation between  
stars and dark matter of $R_c/r_s^d=0.1$ and $0.2$, respectively. As  
shown in Fig.~\ref{fig:dens_evol}, these fits are good over one to two  
decades in surface density, a range comparable to that normally used  
in observations of Local Group dSphs. The panels in  
Figure~\ref{fig:magic} also include the central line-of-sight velocity  
dispersion, $\sigma_0$, measured using stars within one core radius of  
the center.  
  
The main evolutionary trends driven by tides are well defined. {\it  
Tidal effects lower the surface brightness; reduce the core radii;  
increase the King concentration; and reduce the velocity dispersion of  
stars embedded within NFW halos.}  
  
These panels also make clear that the results are highly dependent on  
(i) the pericenter of the orbit; (ii) the number of orbital periods  
completed; as well as (iii) the radial segregation between stars and  
dark matter. The individual impact of these parameters is highly degenerate, so  
that a particular modification in the dSph structure may be achieved  
by trading off the effect of one against another. A certain amount of  
mass loss, for example, may be achieved by completing several orbits  
of moderate pericenter, or by completing a single one with smaller  
$r_{\rm per}$.  
  
Interestingly, the degeneracy between these various effects can be  
collapsed into a single parameter, which may be used to describe the  
structural changes in the stellar component fairly accurately. Such  
parameter may be expressed, for example, as the total mass within the  
(initial) core radius of the unperturbed King model, $M_c\equiv  
M[r<R_c(t=0)]$. This is shown in the panels on the right of  
Fig.~\ref{fig:magic}: irrespective of pericenter, segregation, or  
number of completed orbits, the evolution of the structural parameters  
depends solely on how much (and not {\it how}) the total mass within  
the initial core radius has varied.  
  
This is a remarkable result, especially because our series includes  
extreme cases where the mass within the core radius has decreased by  
up to a factor of $\sim 30$. Such models have lost $99.7\%$ of their  
stars; dropped in surface density by a factor of $\sim 100$; lowered  
their velocity dispersion by a factor of $\sim 4$; and trimmed their  
core radii by a factor of $\sim 3$.

We have fitted a simple empirical formula to the evolution of the  
structural parameters,  
\begin{equation}  
g(x)=\frac{2^\alpha x^\beta}{(1+x)^\alpha},   
\label{eq:gx}  
\end{equation}  
where $x\equiv M_c/M_c(t=0)$ and $g(x)$ represents any of the  
quantities plotted in Fig.~\ref{fig:magic}.  The best fit values of  
$\alpha$ and $\beta$ for each case are listed in Table~\ref{tab:fit},  
and the results are shown by the dashed lines in  
Fig.~\ref{fig:magic}.

\subsection{Evolution of the mass-to-light ratio}  
\label{ssec:evml}  
  
One interesting application of the results presented above is to the  
evolution of the mass-to-light ratio of a dSph, as inferred from the  
velocity dispersion and King-model fit parameters. This is  
traditionally estimated assuming that ``mass follows light'' as  
$\Upsilon_c \simeq 1.447 \sigma_0^2/(\Sigma_0 R_h)$ (Richstone \& Tremaine  
1986). Here $R_h$ is the half-light radius, which may be easily  
computed for a King model from $R_c$ and $c_K$.  
  
\begin{figure}  
\plotone{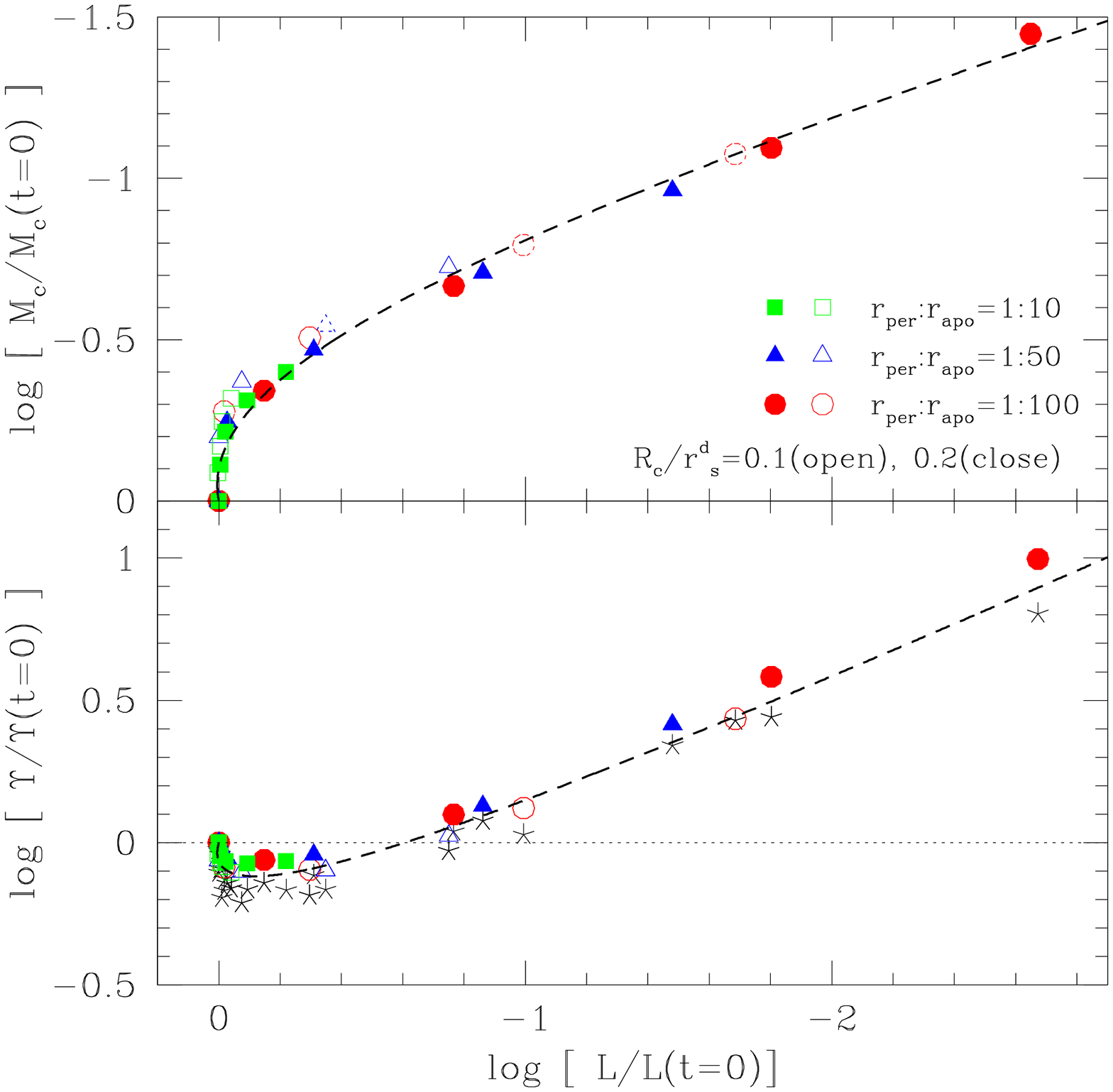}  
\caption{Evolution of the total bound mass within the (initial)  
luminous radius (upper panel) and of the mass-to-light ratio (lower  
panel) as a function of stellar mass loss for different orbits and  
stellar segregations. Starred symbols denote the classical  
mass-to-light ratio estimate $\Upsilon_c \propto \sigma_0^2/(\Sigma_0  
R_h)$ of Richstone \& Tremaine (1986). Other symbols are actual  
simulation measurements (within $R_h$). Note that the mass-to-light  
ratio increases when a dwarf undergoes substantial stripping of  
stars.}  
\label{fig:ml}  
\end{figure}

The result is illustrated in the bottom panel of Fig.~\ref{fig:ml},  
where the starred symbols show $\Upsilon_c$ as a function of the total  
luminosity of the bound stellar component (in units of the initial  
value). This is in good agreement with the evolution of the ``true''  
mass-to-light ratio within the half-light radius $\Upsilon=2  
M^d(<R_h)/L$, which is indicated by the solid symbols in the same  
panel. Two different regimes may be discerned here. One corresponds to  
{\it decreasing} mass-to-light ratio when stellar mass loss is  
moderate: dSphs that have lost less than $2/3$ of their original  
stellar mass see their mass-to-light ratios decrease slightly.  
  
This is because tides actually strip mass from {\it all} radii at all  
times. Indeed, the mass in the inner regions decrease initially as  
tides strip particles on elongated orbits which, although they spend  
most of their time at large radii, they still contribute to the total  
mass in the inner regions. These are preferentially dark matter  
particles, since stars are (by construction) confined to small radii  
by the presence of an outer cutoff in the stellar distribution (i.e.,  
the ``tidal radius'' of the initial King model). As a result, for  
moderate stellar mass loss more dark matter than stars are lost from  
within $R_h$, and $\Upsilon$ decreases. The effect is not large,  
however, and $\Upsilon$ never decreases by more than $\sim 40\%$.  
  
On the other hand, when stars are stripped in earnest the trend is  
reversed; no dark matter is left outside the outer cutoff in the  
stellar component, and most particles are stripped from the luminous  
body of the dwarf. Since the inner dark matter profile is ``cuspy''  
and thus more centrally concentrated than the stars, it is now stars  
that are preferentially lost, and the mass-to-light ratio, $\Upsilon$,  
{\it increases}. This trend dominates the late stages of stripping:  
dwarfs that have lost more than $\sim 90\%$ of their original stars  
are expected to have higher mass-to-light ratios than unperturbed  
systems. A system that has retained only $1\%$ of its stars would have  
a mass-to-light ratio almost a factor of $\sim 10$ higher than its  
initial value. We shall use these results below to assess the possible  
presence of tidal remnants in the Local Group amongst the newly  
discovered ultra-faint dwarfs.  
  
\begin{figure}  
\plotone{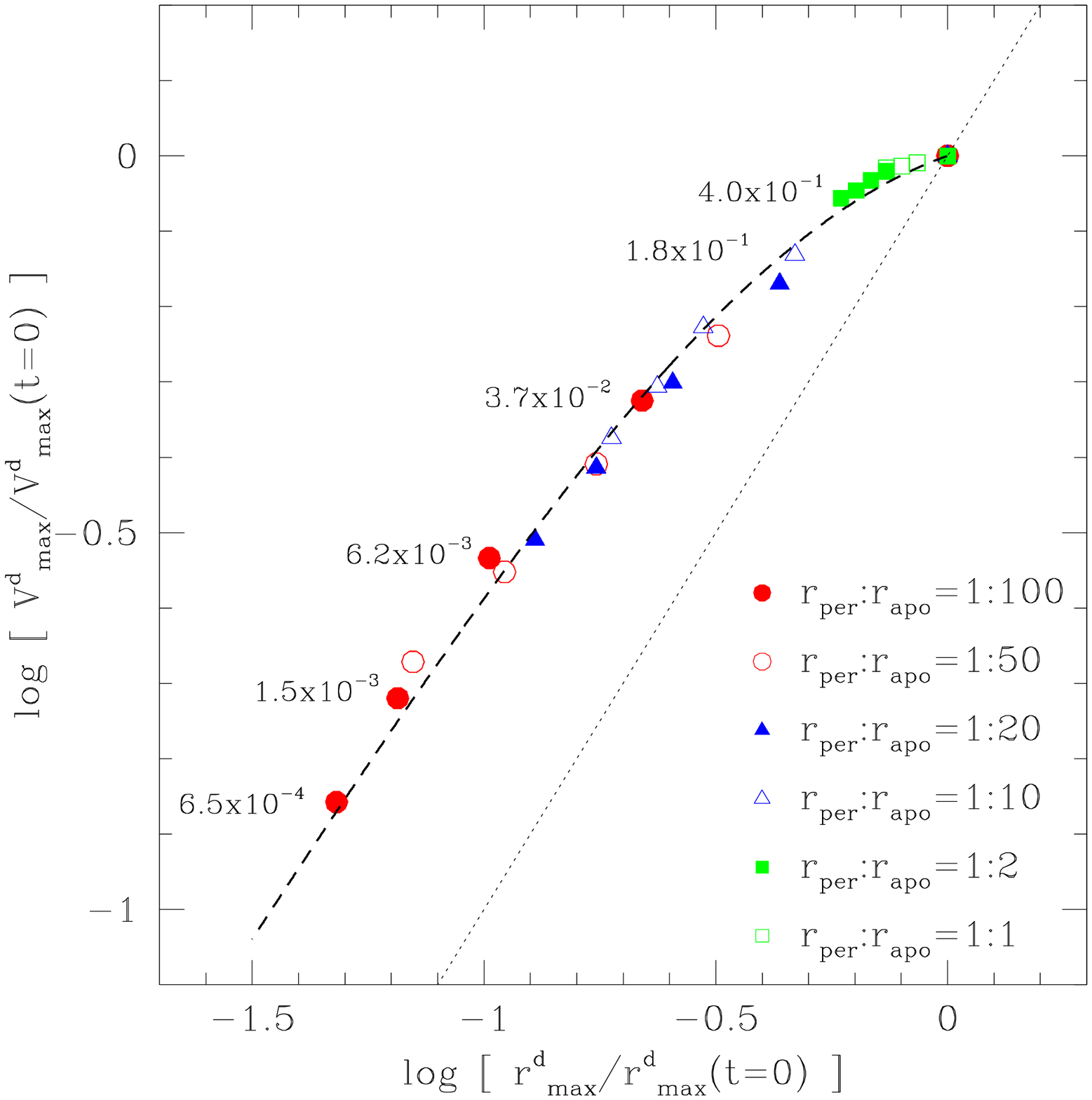}  
\caption{Evolution of the halo peak circular velocity $V_{\rm  
max}\equiv V_c(r_{\rm max})$ and its location, measured at the  
apocenter of the various orbits in our simulation series. The dotted  
line illustrates the roughly $1$:$1$ relation expected between $V_{\rm  
max}$ and $r_{\rm max}$ for unperturbed $\Lambda$CDM halos.  The  
dashed line is our fit to the evolutionary track using eq.~\ref{eq:gx}  
(see Table~\ref{tab:fit}). Labels indicate approximately the total  
bound mass fraction along the curve. }  
\label{fig:vmrm}  
\end{figure}  

\subsection{Evolution of the dark halo}  
\label{ssec:evdmh}  
  
As discussed by PMN, the structural parameters of the stellar  
component of dSphs may be used to estimate the total mass of their  
surrounding dark matter halos. These authors show how $\Sigma_0$ and  
$R_c$ may be used to constrain, under plausible assumptions, the  
physical parameters of the surrounding NFW halo, such as $r_{\rm max}$  
and $V_{\rm max}$. The procedure outlined by PMN, however, assumes  
that halos are well described by an unperturbed NFW profile, an  
assumption that clearly breaks down when the halo has been  
substantially stripped by tides (see, e.g., Hayashi et al 2003,  
Kazantzidis et al 2004). How severely are the estimates of PMN  
affected when tides have been at work?  
  
In order to address this question, we explore first how the parameters  
$r_{\rm max}$ and $V_{\rm max}$ describing the halo evolve as it is  
stripped. This is shown in Fig.~\ref{fig:vmrm}, and indicates, in  
agreement with the earlier results of Hayashi et al (2003), that the  
position of the circular velocity peak is more significantly affected  
than the peak velocity itself. The symbols in Fig.~\ref{fig:vmrm} show  
the results for all our orbits (measured at every apocenter), and show  
that halos follow a well defined path in the $r_{\rm max}$-$V_{\rm  
max}$ plane. In agreement with Hayashi et al (2003), we find that the  
changes in $r_{\rm max}$ and $V_{\rm max}$ depend solely on the total  
{\it amount} of mass lost, and not on the details of {\it how} it was  
stripped. Our most disturbed halo has retained just $0.15\%$ of its  
initial mass; its $r_{\rm max}$ has decreased by a factor of $\sim 30$  
and $V_{\rm max}$ by $\sim 7$.  
  
Note that in the event of large mass loss $r_{\rm max}$ scales almost  
linearly with $V_{\rm max}$, implying that halos evolve in this plane  
along ``tracks'' nearly parallel to the cosmological relation linking  
$r_{\rm max}$ and $V_{\rm max}$ of unperturbed $\Lambda$CDM halos  
(denoted roughly by the dotted line in Fig.~\ref{fig:vmrm}; see NFW or  
PMN for further details).  
  
\begin{figure}  
\plotone{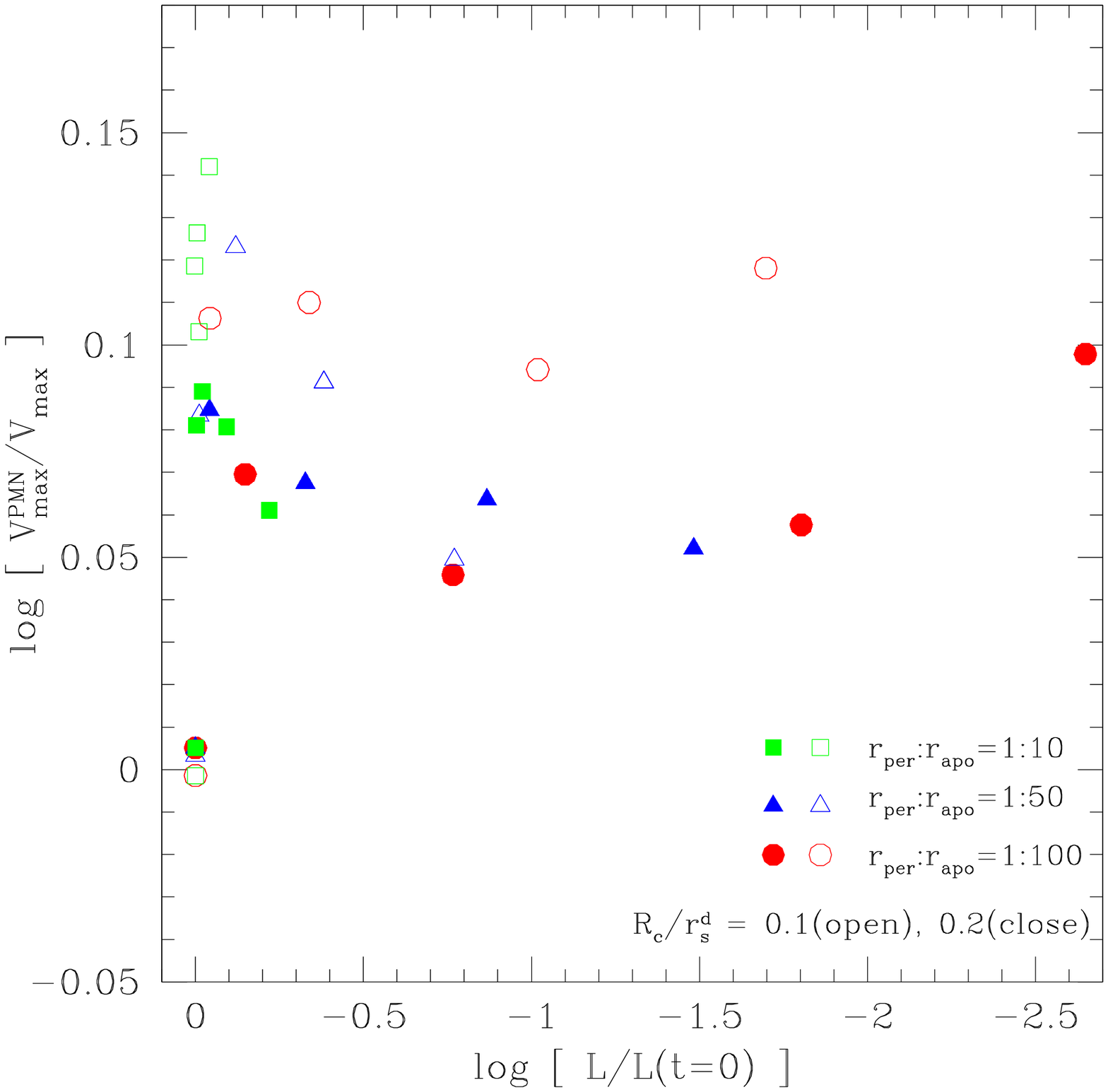}  
\caption{Estimates of $V_{\rm max}$ obtained adopting the procedure of  
Pe\~{n}arrubia, McConnachie \& Navarro (2007, PMN), shown as a  
function of the bound stellar mass (or luminosity). This uses the  
King core radius and the central stellar velocity dispersion to  
estimate the mass within the luminous radius of the dwarf and then  
extrapolates this measurement assuming that the halo follows a  
$\Lambda$CDM NFW profile. The estimates are given in units of the  
``true'' $V_{\rm max}$ measured in our simulation series. Note that  
even in the case of extreme mass loss the PMN procedure yields robust  
estimates of the peak halo circular velocity.}  
\label{fig:estim}  
\end{figure}

As a result, estimating $V_{\rm max}$ by assuming that the halo is an  
unperturbed NFW that follows the cosmological relation between $r_{\rm  
max}$ and $V_{\rm max}$ (as in PMN) results at most in a small  
error. This is shown in Fig.~\ref{fig:estim}, where we show, as a  
function of stripped stellar mass, the error in $V_{\rm max}$ incurred  
by applying the procedure outlined by PMN. Note that even for the most  
extreme case of tidal stripping probed by our simulations, the  
resulting error is less than about $\sim 30\%$. We conclude that the  
velocity dispersion and core radii of dSphs may be used to put strong  
constraints on the maximum circular velocity of its surrounding halo,  
and that the $V_{\rm max}$ estimates presented by PMN are relatively  
insensitive to tidal stripping.  
  
\begin{figure*}  
\plotone{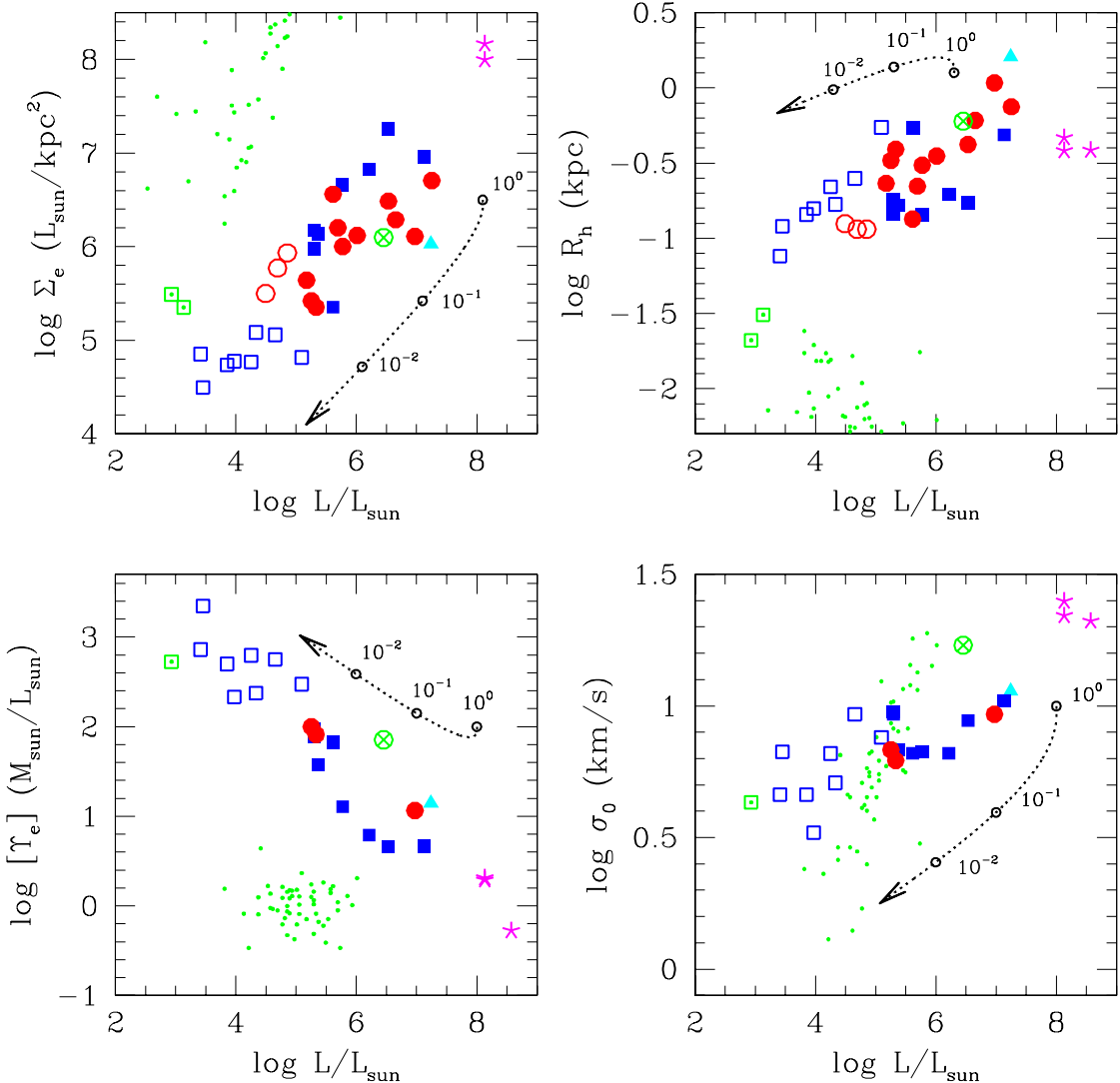}  
\caption{ Local Group dSph structural parameters as a function of  
luminosity in the visual band. Data are compiled from the literature,  
as specified in Table~\ref{tab:obs}. The panels show, respectively,  
the average surface brightness $\Sigma_e\equiv L/(2 \pi R_h^2)$; the  
half-light radius $R_h$; the mass-to-light ratio parameter,  
$\Upsilon_e=5 \sigma_0^2 R_h/GL$; and the central projected velocity  
dispersion. Dwarf spheroidals of the Milky Way are shown with blue  
squares; those of M31 with red circles.  We distinguish between the  
bright, ``classical'' population of dSphs (solid symbols, brighter  
than Draco) and ``ultra-faint'' dSphs (open symbols, fainter than  
Draco). This distinction is arbitrary and is made only to guide the  
discussion in Sec.~\ref{sec:lgdsph}. For illustration, we also show the  
Local Group dwarf ellipticals (magenta stars), the isolated Cetus dSph  
(green crossed circle) and the Milky Way globular clusters (green  
dots). Willman I and Segue I are shown with dotted green squares to  
highlight their intermediate properties between a dSph and a globular  
cluster. Finally, the Sagittarius dSph is shown with a full cyan  
triangle. Dotted lines correspond to the ``tidal evolutionary tracks''  
that a dark matter-dominated dSph would follow as it is gradually  
stripped of stars. Each notch along the track corresponds to the  
conseccutive loss of 9/10 of the stars.}  
\label{fig:tracks}  
\end{figure*}  

\begin{figure*}  
\plotone{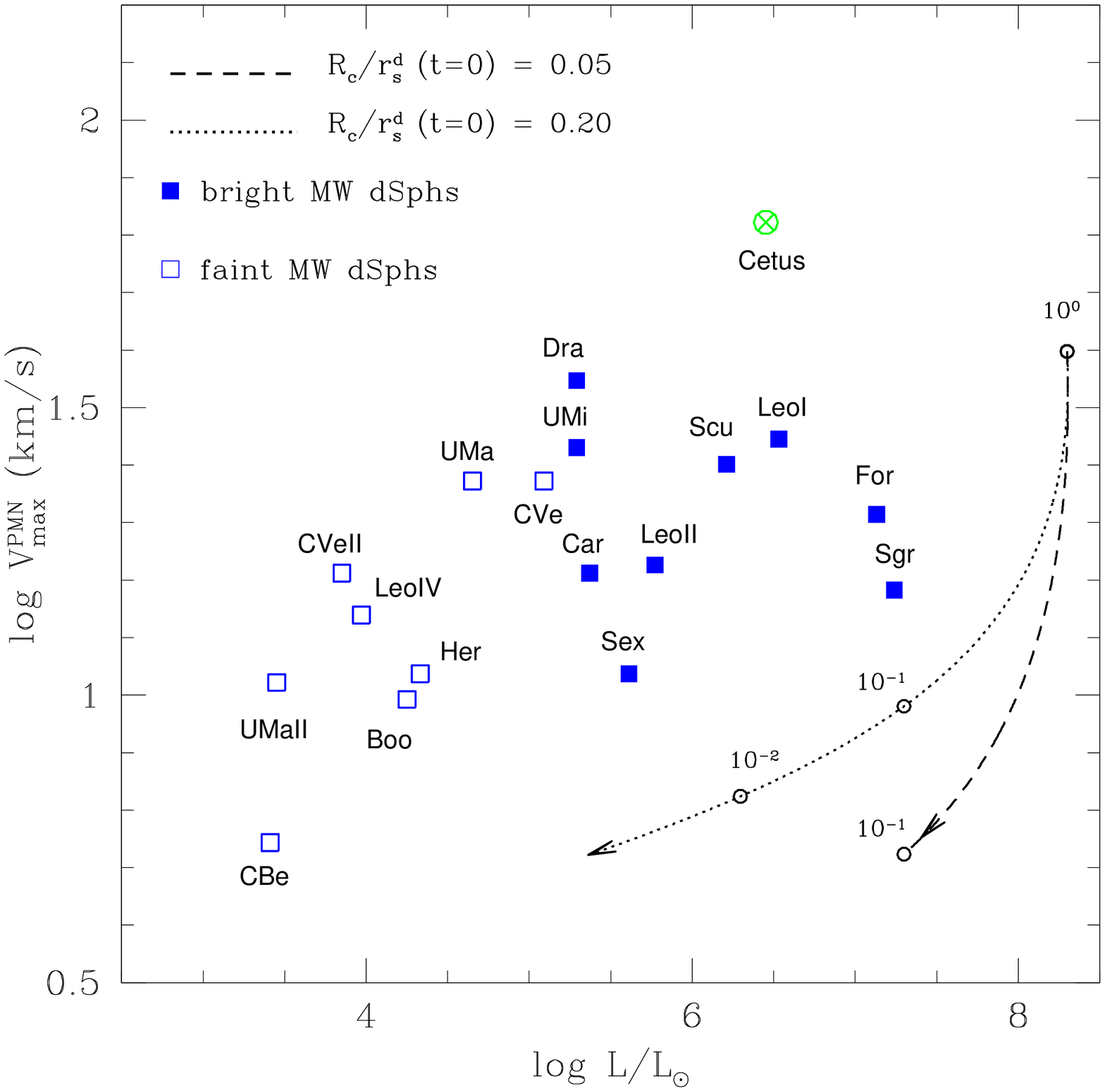}  
\caption{  
Peak halo circular velocity for all LG dSph (estimated following the  
procedure outlined by PMN) as a function of total luminosity. Symbols  
are as in Fig.~\ref{fig:tracks}.  Curves show the expected evolution  
driven by tidal stripping, for two values of the stellar segregation,  
$R_c/r_s^d=0.05$ and $0.2$. The halos of ultra-faint dwarfs (open  
symbols) have on average peak circular velocities comparable to those  
of the ``classical'' dSphs (solid symbols). It is therefore unlikely  
that, as a whole, the faint dwarf population originates from stripping  
once-brighter progenitors like the ``classical'' dSphs. This does not  
exclude the possibility that {\it some} dSphs, and in particular those  
with low estimated $V_{\rm max}$, like CBe, may indeed be tidal  
remnants.  }  
\label{fig:lsigma}  
\end{figure*}  

\section{Application to Local Group dwarfs}  
\label{sec:lgdsph}  
  
The results presented in \S\ref{sec:tidev} allow us to assess the  
effect of tides on the Local Group dSph population and to address some  
of the questions posed in Sec.~\ref{sec:intro}. We begin by  
introducing the latest data available for LG dSphs, compiled from the  
literature, and then discuss the possible role that tides may have  
played on the structure of these objects.  
  
\subsection{The structural properties of Local Group dSphs}  
  
Table~\ref{tab:obs} lists the observational parameters most relevant  
for this discussion, namely the (visual band) luminosity ($L$), the  
half-light radius ($R_h$, measured as the geometric mean of the  
semi-major and minor axes) and the central velocity dispersion  
($\sigma_0$). The table lists all dSphs discovered so far in the Local  
Group, including very recent discoveries. They have been divided by  
environment (MW satellites, M31 satellites and isolated dSphs) and are  
listed in order of decreasing luminosity. We distinguish between  
``classical'' dSphs and ``ultra-faint'' dSphs, as those brighter or  
fainter than Draco, respectively. Also listed are the peculiar systems  
Willman 1 and Segue 1, as well as the isolated Cetus dwarf. Finally,  
Table~\ref{tab:obs} also lists derived quantities, such as the  
effective surface brightness ($\Sigma_e = L/(2\pi R_h^2)$), and the  
mass-to-light ratio parameter. Since for several of the recently discovered dSphs only estimations of their total luminosity and half-light radius are available, we compute the effective mass-to-light ratio as $\Upsilon_e=5 \sigma_0^2 R_h/(G  
L)$. For a King profile, the relationship between $\Upsilon_e$ and $\Upsilon_c$ derived by Richstone \& Tremaine (1986) depends upon the value of the concentration $c_K$, so that for typical values of $c_K=3, 5$ and 10, $\Upsilon_c/\Upsilon_e=1.1, 1.0$ and 0.8, respectively.  
  
Fig.~\ref{fig:tracks} shows these parameters as a function of  
luminosity, using different symbols to highlight the various  
categories alluded to above. The symbols distinguish between Milky Way  
(blue squares) and M31 (red circles) dSphs, globular clusters (green  
dots), the dwarf ellipticals NGC 205, 147 and 185 (magenta stars), as  
well as the peculiar objects Willman I and Segue I (dot-centered  
squares). The isolated dwarph spheroidal Cetus is shown as a green  
crossed circle. (No comparable data are available for Tucana.)  Filled  
symbols correspond to the ``classical'' dSph population of the MW  
and M31; open symbols are used for the population of ultra-faint  
dwarfs.  
  
\subsection{Tidal evolutionary tracks}  
  
In each panel of Fig.~\ref{fig:tracks} the dotted line indicates the  
predicted evolution of a dSph as it is stripped by tides. These  
``tidal tracks'' have been computed using the empirical fits presented  
in Fig.~\ref{fig:magic}. Each notch along the tracks corresponds to  
the consecutive loss of 90\% of the stellar mass (or luminosity). The  
head of the track corresponds to the unperturbed object, whilst the  
arrowhead shows its (relative) position in each panel after losing  
$99.7\%$ of its original stellar mass.  
  
The tracks indicate the likely evolution of {\it any} dSph from its  
current location. Since it is unclear whether dSphs have not {\it  
already} been perturbed by tides, there is some uncertainty as to  
where the evolution of a given dSph would lie along its own ``tidal  
track''.  An unperturbed dSph would evolve along a track constructed  
by translating the head of the track to its current location in each  
of the planes. In cases like Sagittarius, where some mass loss clearly  
has already taken place, the evolution is already underway along its  
``tidal track''.  Despite this uncertainty, a number of general  
inferences may be made in reference to the relative location of  
various systems in the space of parameters shown in  
Fig.~\ref{fig:tracks}.  
  
\subsection{Surface brightness evolution}  
  
The first one concerns the evolution in surface brightness driven by  
tides. As shown by the tidal track in the top-left panel of  
Fig.~\ref{fig:tracks}, the surface brightness drops steeply as a dSph  
is stripped by tides: after losing the first half of its stars the  
surface brightness of a dSph drops by approximately a factor of $\sim  
3$.  This is intriguing, as it may explain why Sagittarius, the  
brightest of all dSphs in the Milky Way, is one of the faintest of the  
``classical'' dSphs in terms of surface brightness. (Fornax, which has  
similar total luminosity as Sagittarius, is almost 10 times brighter  
in $\Sigma_e$.)  
  
Interestingly, at the bright end, the classical dSph populations of  
M31 and MW also differ in their effective surface brightness. As noted  
by McConnachie \& Irwin (2006), the brightest ($L>10^6 \, L_\odot$)  
M31 dSphs are, at given luminosity, about a factor of 2 larger in  
$R_h$ and $\sim 4$ times fainter in surface brightness than their MW  
counterparts (compare, e.g., the brightest solid squares and solid  
circles in the top-left and top-right panels of  
Fig.~\ref{fig:tracks}). With the obvious caveat of small-number  
statistics, it is indeed puzzling that satellites formed around two  
seemingly normal spirals should exhibit such difference in  
structure. Could variations in the importance of tides be responsible  
for this offset?  
  
The fact that tides tend to lower both the surface brightness and the  
half-light radius of dSphs {\it simultaneously} argues against this  
hypothesis. For example, if the lower surface brightness of M31 dwarfs  
(top-left panel) reflects the enhanced effect of tides on that  
population, then their progenitors must have had on average a larger  
half-light radius than at present, differentiating them even further  
from their MW counterparts.  
  
Conversely, one could argue that the smaller size of the MW dSphs  
(top-right panel) might be the result from stronger tidal  
stripping. However, that would imply that their progenitors were on  
average brighter in surface brightness than the present-day M31 dSph  
population, again differentiating these systems even further. We  
conclude, therefore, that the differences in structure between the  
``classical'' MW and M31 dSphs cannot be explained by differences in  
the importance of tides.

\subsection{Tidal tracks and scaling relations}  
  
Another interesting result to note from Fig.~\ref{fig:tracks} is that  
the ``tidal track'' in the $L$--$\Sigma_e$ plane is roughly parallel  
to the relation that holds for Local Group dSphs.  Indeed, a $\sim  
10^6\, L_\odot$ dSph that is stripped of $\sim 99\%$ of its stars is  
expected to become a $\sim 10^4 \, L_\odot$ dSph two decades fainter  
in $\Sigma_e$. This is comparable to the observed trend, and suggests  
the intriguing possibility that Fornax-like or Draco-like systems, for  
example, might be the ancestors of some of the newly discovered  
ultra-faint MW dwarfs  
\footnote{Possible exceptions are Segue I and Willman I, which are  
considerably smaller than the rest of dSph of similar luminosity,  
a result that has led to speculation that these systems might actually  
be disrupted globular clusters rather than dwarf spheroidal  
galaxies. A recent measurement of the velocity dispersion of Willman 1  
by Martin et al. (2007), however, indicates that Willman 1 has a very  
high mass-to-light ratio, implying a significant amount of dark matter  
and confirming its identity as a dwarf galaxy.}  
(shown as open blue squares in Fig.~\ref{fig:tracks})  
  
The plausibility of this suggestion is supported by the few available  
estimates of the dynamical mass-to-light ratio. This is shown by the  
``tidal track'' in the $L$-$\Upsilon_e$ panel of  
Fig.~\ref{fig:tracks}. The track again runs roughly parallel to the  
scaling observed for LG dSphs, suggesting that the extreme  
mass-to-light ratios of the newly-discovered dwarfs (at least those  
for which measurements are available) might in principle be due to  
tidal stripping of once brighter systems.   
  
This would be a radically different interpretation of the origin of  
the main scaling relations of dSphs shown in Fig.~\ref{fig:tracks},  
which is usually ascribed to the mass-dependent efficiency of star  
formation, feedback, and re-ionization in low-mass halos (see, e.g.,  
Dekel \& Woo 2003).   
  
One problem with the tidal interpretation, however, is that the  
$L$-$\sigma_0$ relation is much shallower than one would expect from  
the tidal evolutionary track, as shown in the bottom-right panel of  
Fig.~\ref{fig:tracks}. Indeed, there is little evidence for a  
correlation between luminosity and velocity dispersion for most dSphs,  
except perhaps for a hint of a downturn in the case of the faintest  
dwarfs.  
  
A conservative conclusion, then, is that tides cannot induce the  
full range of observed parameters for dSphs, and that therefore the LG  
dSph population must have been born with a significant range of  
properties. Extreme tidal stripping might be responsible for  
setting the properties of some individual dSphs, but are unlikely to  
explain the scaling relations of the whole dSph population.

\subsection{The halo properties of ultra-faint dSphs}  
  
Support for this conclusion comes from comparing the properties of the  
dark halos of the ultra-faint dwarfs with those of brighter dSphs. If  
the ultra-faint dwarfs were originally massive systems that have lost  
a large fraction of their stellar mass, this would show as a net shift  
in the peak circular velocity, $V_{\rm max}$, between faint and bright  
dSphs. In quantitative terms, systems that have been stripped of  
$99\%$ of their stars should have suffered an even higher loss of dark  
matter, and should have seen their $V_{\rm max}$ reduced by a factor  
of $\sim 3$ or so (see Fig.~\ref{fig:vmrm}).  
  
We explore this in the right-hand panel of Fig.~\ref{fig:lsigma},  
where we show the relation between dSph luminosity and {\it inferred}  
$V_{\rm max}$, computed adopting the procedure of PMN (see  
Sec.~\ref{ssec:evdmh}). This panel shows that, although on average  
ultra-faint dSphs have slightly lower values of $V_{\rm max}$, the  
difference is, again, not as large as one would expect if they were  
tidal remnants of the ``classical'' bright dSph population.  This  
suggests again that it is unlikely that the population of faint dSphs  
descends directly from the bright dwarf spheroidals in the Milky Way,  
and that the observed correlations between the structural properties  
of dSphs are not the result of tidal stripping.  
  
Fig.~\ref{fig:lsigma} also shows that the halos hosting dSphs are  
remarkably similar. Indeed, they differ by less than a factor of $\sim  
3$ in $V_{\rm max}$ despite the fact that their luminous components  
span roughly $4$ {\it decades} in luminosity.  This result lends  
support to scenarios that envision dwarf galaxies as able to form only  
in halos above a certain mass threshold. Such a threshold has been  
suggested by the need to retain gas and to sustain continuing star  
formation despite the effects of feedback from evolving stars and the  
heating from photoionizing radiation (Efstathiou 1992, Bullock,  
Kravtsov \& Weinberg 2000, Somerville 2002, Benson et al. 2002).  
  
In such a scenario, the scarcity of luminous dwarfs around the Milky Way  
may be reconciled with the copious substructure seen in N-body  
simulations of CDM halos (Moore et al 1999, Klypin et al 1999) by  
the fact that massive substructure halos that exceed the threshold are  
relatively scarce (Stoehr et al 2002, Hayashi et al 2003, Kazantzidis  
et al 2004). Indeed, only about $\sim 20$ substructure halos with  
$V_{\rm max}>20$ km/s are expected within the Milky Way halo (see  
PMN's Fig.8). This number is similar to the expected number of dSphs  
with such properties, once the number of newly-discovered dwarfs in  
the SDSS is corrected by the finite sky coverage of the survey  
(Koposov et al. 2007). Our results thus seem to support the main tenet of this  
interpretation: most dwarfs at the extreme faint end of the luminosity  
function inhabit relatively massive halos of mass similar to that  
defined by the threshold.

\section{Summary}  
\label{sec:sum}  
  
We have used N-body simulations to study the dynamical evolution of  
dwarf spheroidal galaxies (dSphs) driven by galactic tides. We assume  
a cosmologically motivated scenario where dSphs are modelled as  
dark-matter dominated systems whose stellar component may be  
approximated as King models embedded within NFW halos. These systems  
are in eccentric orbits in the potential of a much more massive halo,  
also modelled using an NFW profile.  Under these circumstances, tides  
operate nearly impulsively at pericenter and the dwarf relaxes quickly  
afterwards. We focus our analysis on the equilibrium structure reached  
by a dwarf following repeated episodes of mass loss. Our main  
conclusions may be summarized as follows.  
  
\begin{itemize}  
  
\item Our models show that NFW-embedded King models are {\it  
extraordinarily resilient} to tides; the density profile of the  
stellar component still resembles a King model even after losing more  
than 99\% of its stars.  
  
\item The King-model structural parameters evolve as tides strip the  
galaxy. The stellar velocity dispersion, $\sigma_0$, central surface  
brightness, $\Sigma_0$, and core radius, $R_c$, decrease  
monotonically, in a manner controlled solely by the total amount of  
mass lost from within the luminous radius.  
  
\item The core radius, $R_c$, is the parameter least affected by  
tides: after losing 99\% of the stars, $R_c$ decreases by only a  
factor of $\sim 2$. Thus, even in the event of extreme mass loss the  
core radius is a robust measure of the original size of the system.  
  
\item After substantial mass loss, tides tend to make dSphs {\it more  
dark-matter dominated}. This is because the tightly bound central dark  
matter ``cusp'' is more resilient to disruption than the comparatively  
more loosely bound ``cored'' King profile. Tidal effects may therefore  
help to explain the extremely large mass-to-light ratios of some of  
the newly-discovered ultra-faint Milky Way dwarfs.   
  
  
\item Although tides may induce changes in luminosity, surface  
brightness and mass-to-light ratio, that mimic the observed  
correlations, the weak (if any) luminosity dependence of the velocity  
dispersion implies that tides cannot be responsible for the scaling  
relations and full range of structural parameters of the Local Group  
dSphs.  
  
\item The above conclusion is supported by the large peak circular  
velocities ($V_{\rm max}$) inferred for the ultra-faint dSphs. With  
few exceptions, these are comparable to those of the bright dSphs,  
implying that it is unlikely that they are the true descendents of the  
present-day population of bright dSphs.  
  
\item Despite the large range in luminosity spanned by dSphs, they  
seem to inhabit halos with a surprisingly narrow range of peak  
circular velocity (or mass). This provides support for a scenario  
where dwarf galaxies are only able to form in halos above a certain  
mass threshold.  
  
\end{itemize}  
  
This is clearly a fast moving field where the tentative conclusions  
listed above are likely to be superseded as new data becomes  
available. In particular, spatially-resolved dynamical data for the  
newly-discovered dwarfs, as well as kinematic data for M31 dSphs  
should prove invaluable in order to constrain further current  
theoretical models of the formation of dSphs; the faintest galaxies in  
the Local Universe and certainly a puzzling collection of  
extragalactic elves.  
  
\vskip1cm This work has been supported by various grants to JFN from  
Canada's NSERC. JFN acknowledges useful discussions with Simon White,  
as well as the hospitality of the Max-Planck Institute for  
Astrophysics in Garching bei Muenchen, Germany. Stelios Kazantzidis is  
kindly thanked for providing the code used to generate the spherical  
NFW equilibrium N-body models.  
   
{}

\newpage  
\begin{table}   
\begin{center}   
\caption{Orbital parameters of our models}  
\begin{tabular}{l  c  c c c } \hline \hline   
$r_{\rm per}$:$r_{\rm apo}$ & $r_{\rm per}$ & $T_r$ & $r_{\rm tid}$ & $M^d(<r_{\rm tid})$ \\   
 & [kpc] & [Gyr] & [$r_s^d$] & [$M^d_{\rm tot}$] \\   
\hline  
1:1   & 180.0   &  $\infty$  & 9.75 & $6.4\times 10^{-1}$ \\  
1:2   & 90.0    & 4.77          & 4.65  & $3.9\times 10^{-1}$ \\  
1:10  & 18.0   & 3.36	      & 0.70  & $5.2\times 10^{-2}$ \\  
1:20  & 9.0    & 3.08         & 0.28  & $1.2\times 10^{-2}$ \\  
1:50  & 3.6    & 2.94         & 0.08  & $1.3\times 10^{-3}$ \\  
1:100 & 1.8    & 2.80         & 0.03  & $2.7\times 10^{-4}$ \\  
\hline    
\newline   
\newline   
\end{tabular}\label{tab:mod}   
\end{center}   
\end{table}   
  
\begin{table}   
\begin{center}   
\caption{Empirical fit parameters to the tidal evolutionary tracks  
(eq.~\ref{eq:gx})}  
\begin{tabular}{l c c c c c c c} \hline \hline   
 &   
$V_{\rm max}$ &  
$\Sigma_0$  &  
$R_c$ &  
$c_K$ &  
$\sigma_0$ &  
$L$  &   
$\Upsilon$ \\   
 &   
$(r_{\rm max})$ &  
$(M_c)$  &  
$(M_c)$ &  
$(M_c)$ &  
$(M_c)$ &  
$(M_c)$  &   
$(M_c)$ \\   
\hline  
$\alpha$ & 1.55   & 2.70  & 1.50  & 1.75  & 0.00  & 7.00 & -4.80 \\   
$\beta$  & 0.99   & 2.00  & 0.65 & 0.00   & 0.50  & 3.30 & -1.60 \\   
\hline    
\newline   
\newline   
\end{tabular}\label{tab:fit}   
\end{center}   
\end{table}   
  
\begin{table}   
\begin{center}   
\caption{Structural Parameters of Local Group Satellites}  
\begin{tabular}{l c c c c c c c} \hline \hline   
Galaxy &   
$\log_{10}L$ &  
$R_h$ &   
$\log_{10}\Sigma_e$  &   
$\sigma_0$ &  
$\Upsilon_e$ &  
Symbol  &  
Refs.\\   
 &   
[$L_\odot$] &  
[kpc] &   
[$M_\odot$ kpc$^{-2}$]  &   
[km/s] &   
$M_\odot/L_\odot$ &  
[Fig.~\ref{fig:tracks}]  
  &  
\\ \hline  
 Sagittarius  & 7.24 &     1.61&      6.03 &     11.4 & 14.0 & solid triangle & (1)\\  
 Fornax       & 7.13 &    0.49 &      6.96 &     10.5 & 4.6  & solid square & (2),(3)\\  
 Leo I        & 6.53 &    0.17 &      7.26 &     8.8 & 4.6  & ``& (2),(3)\\  
 Sculptor     & 6.21 &    0.20 &      6.83 &     6.6 & 6.14 & ``& (2),(3)\\  
 Leo II       & 5.77 &    0.14 &      6.66 &     6.7 & 12.7 &``& (2),(3)\\  
 Sextans      & 5.61 &    0.54 &      5.35 &     6.6 & 66.7 &``&  (2),(3)\\  
 Carina       & 5.37 &    0.17 &      6.13 &     6.8 & 38.0 &``&  (2),(3)\\  
 Ursa Minor   & 5.29 &    0.18 &      5.98 &     9.3 & 93.1 &``&  (2),(3)\\  
 Draco        & 5.29 &    0.15 &      6.17 &     9.5 & 77.9 & ``&  (2),(3) \\   
\hline  
Canes Venatici& 5.09 &    0.55 &      4.82 &     7.6 & 297.9 & open square & (6,12,15)\\  
Ursa Major    & 4.65 &    0.25 &      5.06 &     9.3 & 561.6      &  ``& (5,13,14,15)\\   
Hercules      & 4.33 &    0.17 &      5.08 &     5.1 & 237.1      & ``& (9,15)\\  
Bootes        & 4.25 &    0.22 &      4.77 &     6.6 & 625.2      & ``& (8,11,14)\\  
Leo IV        & 3.97 &    0.16 &      4.77 &     3.3 & 213.9      & ``& (9,15)\\  
Canes Venatici II & 3.85&   0.14 &      4.74 &   4.6 & 499.3      & ``&(9,15)\\  
Ursa Major II & 3.45 &    0.12 &      4.49 &     6.7 & 2217.0     & ``&(7,14,15)\\  
Coma Berenices & 3.41 &    0.08 &      4.85 &     4.6 & 725.8       & ``&(9,15)\\   
\hline   
\hline   
And VII       &7.25 &     0.75 &       6.71 &       ---& --- & solid circle & (16)\\  
And II        &6.97 &     1.10 &       6.11 &       9.3&  11.6 & ``& (16)\\  
And I         &6.65 &     0.61 &       6.29 &       ---& ---& ``& (16)\\  
And VI        &6.53 &     0.42 &       6.49 &       ---& ---&``& (16)\\  
And III       &6.01 &     0.35 &       6.12 &       ---& ---&``& (16)\\  
And V         &5.77 &     0.31 &       6.00 &       ---& ---&``& (16)\\   
And XV        &5.69 &     0.22 &       6.20 &       ---&---& `` & (20)\\  
And XVI       &5.61 &     0.13 &       6.56 &       ---&---&`` & (20)\\  
And XIV       &5.33 &     0.39 &       5.35 &       6.2& 81.3 &`` & (21)\\  
And IX        &5.25 &     0.33 &       5.42 &      6.8& 99.2 &`` & (17,22)\\  
And X         &5.17 &     0.23 &       5.64 &       ---& ---&`` & (18)\\  
\hline  
And XI        &4.85 &     0.11 &       5.93 &       ---& ---&open circle & (19)\\  
And XIII      &4.69 &     0.12 &       5.77 &       ---& ---&`` & (19)\\  
And XII       &4.49 &     0.13 &       5.50 &       ---& ---&`` & (19)\\   
\hline  
Segue I       &3.13 &     0.03 &       5.35 &      ---& ---&dotted square & (9)\\  
Willman I     &2.93 &     0.02 &       5.49 &      4.3&  529.2&`` & (4,14)\\   
\hline  
Cetus         &6.45 &     0.60 &       6.01 &      17.0& 71.4 & crossed circle & (16,24)\\  
\hline    
\newline   
\newline   
\end{tabular}\label{tab:obs}   
\tablerefs{(1) Majewski et al. (2003); (2) Mateo (1998); (3) Irwin \&  
Hatzidimitriou (1995); (4) Willman (2005a); (5) Willman et  
al. (2005b); (6) Zucker et al. (2006a); (7) Zucker et al. (2006b); (8)  
Belokurov et al. (2006b); (9) Belokurov et al. (2007); (10) Walsh et  
al. (2007); (11) Mu\~noz et al. (2006); (12) Ibata et al. (2006); (13)  
Kleyna et al. (2005); (14) Martin et al. (2007); (15) Simon \& Geha  
(2007); (16) McConnachie\& Irwin (2006); (17) Harbeck et al. (2005);  
(18) Zucker et al. (2007); (19) Martin et al. (2006); (20) Ibata et  
al. (2007); (21) Majewski et al. (2007); (22) Chapman et al. (2005);  
(23) Coleman et al. (2007); (24) Lewis et al. (2007) }  
\end{center}   
\end{table}

\end{document}